\newif\ifhasbib

\hasbibtrue

\documentclass[twocolumn]{aastex62}

\shorttitle{Exoplanet Population Synthesis in the Era of Kepler}
\shortauthors{Mulders et al.}

\usepackage{natbib,graphicx,xspace,amsmath,footnote}
\usepackage{afterpage}

\newcommand{\epos}{\texttt{epos}\xspace}

\newcommand{\kepler}{\textit{Kepler}\xspace}
\newcommand{\bern}{\textit{Bern}\xspace}

\newcommand{\setlabelscale}{\setlength{\unitlength}{0.45\linewidth}}
\newcommand{\setlabelscalesingle}{\setlength{\unitlength}{0.45\textwidth}}
\newcommand{\putlabel}[3][1]{\put(-#1,#2){\color{brown}{\bf #3})}}

\newcommand{\Panel}[1]{Panel {\color{brown}{\bf #1})}\xspace}
\newcommand{\panel}[1]{{\color{brown}{\bf #1})}\xspace}
\newcommand{\panelp}[1]{({\color{brown}{\bf #1}})\xspace}

\newcommand{\sfig}[2]{Fig. \ref{f:#1}{\color{brown}{\bf #2}}}
\newcommand{\sfigp}[2]{(Fig. \ref{f:#1}{\color{brown}{\bf #2}})}

\newcommand{\rfigp}[1]{(Fig. \ref{f:#1})}
\newcommand{\sfigure}[2]{Figure \ref{f:#1}{\color{brown}{\bf #2}}}
\newcommand{\rfigure}[1]{Figure \ref{f:#1}}

\newcommand{\dpr}{\ensuremath{\mathcal{P}}\xspace}

\definecolor{tropicalrainforest}{rgb}{0.0, 0.46, 0.37}

\usepackage{xcolor, fontawesome, microtype}
\definecolor{twitterblue}{RGB}{64,153,255}
\definecolor{linkcolor}{rgb}{0.1216,0.4667,0.7059}
\newcommand{\twitter}[1]{\href{https://twitter.com/#1}{\textcolor{twitterblue}{\faTwitter}\,\tt \textcolor{twitterblue}{@#1}}}

\defcitealias{2018AJ....156...24M}{Paper I}

\begin{document}

\title{The Exoplanet Population Observation Simulator. II - Population Synthesis in the Era of Kepler}

\correspondingauthor{Gijs D. Mulders}
\email{mulders@uchicago.edu}


\author{Gijs D. Mulders}
\affil{Department of the Geophysical Sciences, The University of Chicago, 5734 South Ellis Avenue, Chicago, IL 60637 \twitter{GijsMulders}}
\affil{Earths in Other Solar Systems Team, NASA Nexus for Exoplanet System Science}

\author{Christoph Mordasini}
\affil{Physikalisches Institut, Universit\"{a}t Bern, Gesellschaftstrasse 6, 3012 Bern, Switzerland}

\author{Ilaria Pascucci}
\affil{Lunar and Planetary Laboratory, The University of Arizona, Tucson, AZ 85721, USA}
\affil{Earths in Other Solar Systems Team, NASA Nexus for Exoplanet System Science}
\author{Fred J. Ciesla}
\affil{Department of the Geophysical Sciences, The University of Chicago, 5734 South Ellis Avenue, Chicago, IL 60637}
\affil{Earths in Other Solar Systems Team, NASA Nexus for Exoplanet System Science}

\author{Alexandre Emsenhuber}
\affil{Lunar and Planetary Laboratory, The University of Arizona, Tucson, AZ 85721, USA}
\affil{Physikalisches Institut, Universit\"{a}t Bern, Gesellschaftstrasse 6, 3012 Bern, Switzerland}

\author{D\'aniel Apai}
\affil{Department of Astronomy, The University of Arizona, Tucson, AZ 85721, USA}
\affil{Lunar and Planetary Laboratory, The University of Arizona, Tucson, AZ 85721, USA}
\affil{Earths in Other Solar Systems Team, NASA Nexus for Exoplanet System Science}

\begin{abstract}
The collection of planetary system properties derived from large surveys such as \textit{Kepler} provides critical constraints on planet formation and evolution. 
These constraints can only be applied to planet formation models, however, if the observational biases and selection effects are properly accounted for.
Here we show how \epos, the Exoplanet Population Observation Simulator, can be used to constrain planet formation models 
by comparing the \textit{Bern} planet population synthesis models to the \kepler exoplanetary systems.
We compile a series of diagnostics, based on occurrence rates of different classes of planets and the architectures of multi-planet systems within 1 au, that can be used as benchmarks for future and current modeling efforts.
Overall, we find that a model with 100 seed planetary cores per protoplanetary disk provides a reasonable match to most diagnostics.
Based on these diagnostics we identify physical properties and processes that would result in the \textit{Bern} model more closely matching the known planetary systems. These are: 
moving the planet trap at the inner disk edge outward;
increasing the formation efficiency of mini-Neptunes;
and reducing the fraction of stars that form observable planets;
We conclude with an outlook on the composition of planets in the habitable zone, and highlight that the majority of simulated planets smaller than $1.7$ Earth radii in this zone are predicted to have substantial hydrogen atmospheres.\\ 

The software used in this paper is available online for public scrutiny at \href{https://github.com/GijsMulders/epos}{\color{linkcolor}\faGithub\,\url{https://github.com/GijsMulders/epos}}.
\end{abstract}


\section{Introduction}
Exoplanet surveys have revealed a diverse population of planets with a wide range of sizes, orbital period, and system architectures.
Their origins, however, are not well constrained observationally. 
While follow-up observations can provide important clues to the chemical composition of planets and their atmospheres, such observations are limited to a small and biased set of predominantly large and hot exoplanets.
Thus, studying the formation of planets and their emergent ensemble properties are crucial for understanding the origins of the entire exoplanet population \citep[e.g.][]{2019BAAS...51c.475A}.

The diversity in exoplanet properties likely results in part from a range of initial protoplanetary disk properties and evolutionary pathways. Further, the late stages of planet formation are inherently stochastic, with simulations with similar initial conditions producing a wide range of planets and planetary systems \citep[e.g.][]{2009Icar..203..644R,2014E&PSL.392...28F}. Thus linking the properties of the observed exoplanet population to a formation history can be quite challenging and requires a robust statistical analysis comparing model outcomes to the available data.

Population synthesis models have been used to understand how the interplay between initial disk properties and various physical processes shapes the planetary systems that we ultimately observe (see \citealt{2018haex.bookE.143M} for a recent review). The processes accounted for in these models include, but are not limited to, the evolution and dispersal of protoplanetary disks, the growth of planetary cores, the accretion and loss of gaseous envelopes, the migration of planets, and gravitational interactions between protoplanets. These models then make quantitative predictions about the physical properties and chemical compositions of planets, and could ultimately be used to asses the relative importance of different planet formation channels.

For example, the majority of exoplanets found by \kepler have substantial hydrogen atmospheres \citep[e.g.][]{2013ApJ...772...74W} that must have accreted in a gas-rich environment --- in contrast with the gas-poor assembly of the solar system terrestrial planets. 
The planet population identified by \kepler extend close to the habitable zones of Sun-like stars and can be used to estimate $\eta_\oplus$, the fraction of stars with Earth-sized planets in the habitable zone. Studies that extrapolate the planet occurrence rates from regions where those statistics are reliable typically find $\eta_\oplus \sim 30\%$ \citep{2015ApJ...809....8B,2018AJ....156...24M}. Studies that estimate $\eta_\oplus$ directly from planet candidates in the habitable zone find that a wide range of rates is still possible due to the low number of reliable detections \citep{2019AJ....158..109H}.

If exo-Earths do indeed follow different formation pathways than our Solar System, then it is likely that their bulk compositions would vary significantly from those found around the Sun.  Indeed, the need for different formation pathways is seen in looking at the planets around M dwarfs; Traditional Solar System-based models predict that large numbers of low-mass planets would be found around such stars \citep{2007ApJ...669..606R,2015ApJ...804....9C} while surveys indicate that such systems contain much more massive planets, more so than around the higher mass stars \citep{2015ApJ...807...45D,2015ApJ...814..130M}.  The TRAPPIST-1 system \citep{2017Natur.542..456G} is a clear example as it hosts 7 Earth-mass planets inside 0.07 au despite being ~10\% the mass of the Sun. These planets are thought to have water-rich compositions \citep{2018RNAAS...2c.116U}, possibly indicating that inward migration was a critical factor in their formation \citep[e.g.][]{2017A&A...604A...1O}.

While the importance of different formation pathways for exoplanets is evident, the relative importance of various processes in defining these pathways of the observed exoplanet population is less clear. 
Planet population synthesis models make quantitative predictions about the occurrence rates of planets and planetary systems as different processes are introduced to the simulations. Hence, statistical comparisons of model results with observed exoplanet populations can be used to calibrate the relative importance of these formation channels. 

Early planet population synthesis models were mainly focussed on the giant planet populations that were known at the time \citep{2004ApJ...604..388I}, and often simulated the formation and growth of one planet per disk \citep[e.g.][]{2009A&A...501.1139M}. These models predicted a dearth of close-in super-earths, but such planets were later found to be abundant in more sensitive exoplanet surveys \citep{2010Sci...330..653H,2012ApJS..201...15H}. 
With the inclusion of multi-planet interactions and type-I migration, the Bern planet population synthesis models (\citealt{2018haex.bookE.143M}; Emsenhuber et al. in prep) are now capable of producing planets across the entire range in planet size and orbital period where they are observed with \kepler \citep{2013A&A...558A.109A,2018ApJ...853..163J}. 
A crucial next step is to evaluate whether the predicted planet populations also match the statistical properties of observed exoplanets.

In this paper, we use \epos, the Exoplanet Population Observation Simulator \citep{2018AJ....156...24M}, to evaluate the ability of the latest Bern planet population synthesis model \citep{2018haex.bookE.143M} to produce planets and planetary systems that are consistent with the \kepler exoplanet population. 
We assess the relative success of the Bern model in reproducing current exoplanet statistics, and identify areas where the predicted planet populations fall short of doing so. 
We conclude by identifying parameters and processes that should be investigated further that likely would result in the models producing planets and system architectures more consistent with the known exoplanetary systems.

This study serves as a template for future comparisons between planet formation models and exoplanet surveys. As such,
the software to produce the results and figures in this paper is publicly available as a \texttt{Python} repository at \url{https://github.com/GijsMulders/epos} or with \texttt{pip install epospy}.

%
%

\begin{figure*}
    \centering
    \includegraphics[width=\linewidth]{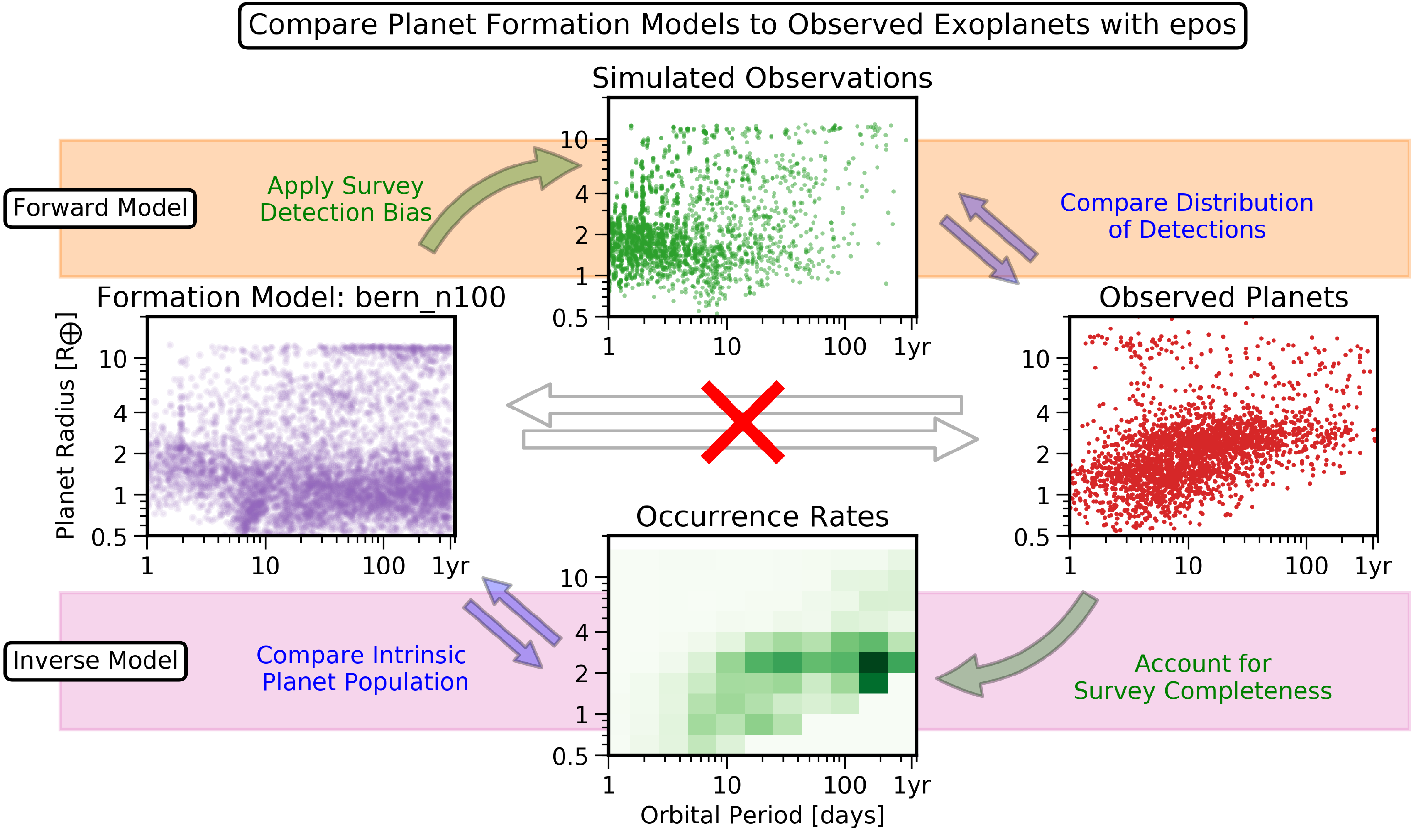}    
    \caption{Outline of the forward modeling approach employed in \epos compared with the occurrence rate calculations (inverse model) used for comparison in this paper, e.g. Fig. \ref{f:occrates}. 
    }
    \label{f:arrows}
\end{figure*}

\section{The Exoplanet Population Observation Simulator}
The Exoplanet Population Observation Simulator, \epos\footnote{\url{https://github.com/GijsMulders/epos}}, is a software package designed to simulate survey observations of synthetic exoplanet populations. By taking into account detection biases and survey completeness, models of the planet population can be constrained from or compared to an observed population of exoplanets (Figure. \ref{f:arrows})

In \citealt{2018AJ....156...24M} (hereafter \citetalias{2018AJ....156...24M}) we demonstrated the basic approach of \epos \texttt{v1.0} \citep{eposv1}. In that paper we used parametric distributions of planet properties to test the methodology and make inferences about the underlying population of \kepler exoplanets using \texttt{emcee} \citep{2013PASP..125..306F}. In this paper, we will expand \epos with simulated planetary systems from the \bern planet population synthesis model. In this section, we will briefly describe how \epos works and how we implement simulated observations of planet formation model outcomes.

\subsection{Basic Approach}\label{s:basic}
The Exoplanet Population Observation Simulator is a software package designed to take into account detection biases in exoplanet surveys. The basic approach has been outlined in \citetalias{2018AJ....156...24M} and summarized here, consisting of 5 steps.

\begin{description}
\item [Step 1] Define a distribution of planetary systems. \citetalias{2018AJ....156...24M} used analytic forms for planet occurrence rates and planetary architectures. Here, we will use the orbital architectures of the \textit{Bern} planet population synthesis model. 
\item [Step 2] From this derive a transiting planet population using a Monte Carlo approach by assigning a random orientation to each system and evaluating which planets transit their host stars.
\item [Step 3] Determine which of the transiting planets would be detected by \kepler, accounting for the geometric transit probability and survey completeness. 
\item [Step 4] Compare the detectable planet population with exoplanet survey data using a set of summary statistics of planetary system properties.
\item [Step 5] In \citetalias{2018AJ....156...24M} we repeated steps 1-4 until the simulated detectable planet population matches the observed planet population to constrain the intrinsic distribution of planetary systems. Instead, we discuss what parameters and processes may need to be further investigated to yield better matches between modeled and real exoplanet systems.
\end{description}

We do not specifically take into account variations in planet occurrence with stellar properties such as mass, metallicity, and potentially age. Instead, we assume that variations in planet properties will average out across the survey, and that the observed planetary system properties are representative of a $5$ Gyr old, solar mass, solar metallicity star.

\subsection{Implementation of Planet Formation Mode}
We implement a ``planet formation mode'' into version \texttt{v2.0} \citep{eposv2} of \epos\footnote{\url{https://github.com/GijsMulders/epos}} to simulate observations of a planet formation model. We adjust steps 1,4, and 5 in as follows. 

In step 1, we simulate a survey of $n_\star$ stars, where a fraction $\eta_s$ has a planetary system. Each planetary system is randomly drawn from a population synthesis model, and has properties $\{R_p, P, \delta i\}_k$, where: 
$R_p$ is the planet radius; 
$P$ is the orbital period;
$\delta i$ is the inclination of the planet with respect to the system inclination $i$;
and $k$ is an index for each planet in the system counted from inside out. 
The system inclination, $i$, and longitude of ascending node for each planet's orbit $\Omega_k$, are generated using a Monte Carlo approach as described in \citetalias{2018AJ....156...24M}. 

We then carry out a synthetic survey by determining which planets are transiting and detectable as in steps 2 and 3 to generate an observable distribution. 
We describe the comparison with observations, the equivalent of step 4, in Sections \ref{s:occ} and \ref{s:multi}. We do not perform the MCMC fitting (step 5) to refine parameters in the models, but highlight which features and processes could be modified to improve the match between the synthetic surveys and that of Kepler.

\subsection{Kepler Survey Completeness}
The survey completeness is defined as the probability that the \kepler pipeline detects a transiting planet of the same size and orbital period and classifies it as a reliable planet candidate (e.g. vetting). The detection efficiency of the \kepler pipeline for each star is calculated with \texttt{KeplerPORTs} \citep{2017ksci.rept...19B}. We use the stellar radii and giant star classification from \textit{Gaia} \citep{2018ApJ...866...99B} which yields ($n=122,682$) main sequence stars,  mainly of spectral types F, G, and K. The survey detection efficiency is then the average detection efficiency of all main sequence stars. While it is possible to select a narrower spectral type range to select the stars most similar to the sun, in practice this does not affect the survey detection efficiency because the average star in the \kepler survey is already of similar mass and radius to the sun. To optimize the number of detected planets in the sample, we choose the larger stellar sample, which is representative of sun-like stars.

The planet candidate vetting efficiency is calculated based on the procedure in \cite{2018ApJS..235...38T} assuming a planet candidate is a reliable detection if the disposition score is above 0.9. We parametrize the vetting efficiency as a double broken power-law in period and radius as described in \citetalias{2018AJ....156...24M}. 

We do not explicitly account for the effects of stellar multiplicity or astrophysical false positives. The \kepler planet host sample has received extensive follow-up observations and analysis \citep{2017AJ....153...71F,2018ApJ...861..149F}. As a result the false positive probability of planet candidates in the \texttt{DR25} catalog is below $1\%$ \citep{2016ApJ...822...86M} and the impact of stellar multiplicity on planet occurrence rates is minimal \citep[e.g.][]{2018AJ....156...83Z}.

\subsection{Parametric Distributions}\label{s:para}
Here we summarize the parametric distributions from \citetalias{2018AJ....156...24M} that we use for the comparison with the \bern planet population synthesis model in Sections \ref{s:bern} and to guide the interpretation of the simulated observables in Section \ref{s:multi}. 

The parametric distributions define five independent distributions of planet properties: the orbital period, $P$; the planet Radius, $R_p$; the mutual inclination\footnote{We refer to the inclination between a planet and the system inclination as \textit{mutual} inclination to discriminate between the inclination of a planet with respect to the observer.} of planets in multi-planet systems, $\delta i$; the period ratio between planets, $\mathcal{P}$; and the period of the innermost planet, $P_0$. Each distribution was assumed to be independent of the other. 
\begin{itemize}
\item The orbital period distribution is a broken power-law with a break at $P\sim12$ days. Interior to the break, the occurrence rate of sub-Neptunes rapidly decreases.
\item The planet radius distribution is also a broken power-law with a break at $R_p\sim3 R_\oplus$. Above the break, planets become rarer with increasing size.
\item The mutual inclination distribution is a Rayleigh distribution with a mode at $\delta i=2\degr$.
\item The period ratio distribution is generated from a log-normal distribution in dimensionless spacing. The median period ratio is $\mathcal{P}=2.1$
\item The period distribution of the innermost planet is a peaked distribution described by a broken power-law with a break at $P\sim12$ days. 
\end{itemize}

The purpose of the comparisons with parametric distributions is to develop an intuitive understanding of the major differences between the model outcomes and the intrinsic planet population that may not be immediately apparent from the direct observables. For example, the mutual inclination distribution is not a direct observable, but has to be derived from multi-planet frequencies using a set of assumptions about planet architectures. 
Of course, some of those assumptions may not be justified by the observations or the planet formation model, and may skew the derived parameters.
For example, as \cite{2018ApJ...860..101Z} have recently pointed out, a different set of assumptions may lead to very different conclusions on the mutual inclination distribution. 
A similar case can be made for the period and radius distributions, which are predicted to be covariant due to the sculpting effect of photo-evaporation and which may bias estimates of $\eta_\oplus$ \citep{2018MNRAS.479.5303L,2019ApJ...883L..15P}. 
Therefore, we will use the parametric distributions for an initial comparison with the models but also perform a synthetic survey to make a direct comparison between the simulated systems and the \kepler exoplanets in Section \ref{s:multi}.

%
%

\begin{figure*}
    \centering
    \includegraphics[width=0.45\linewidth]{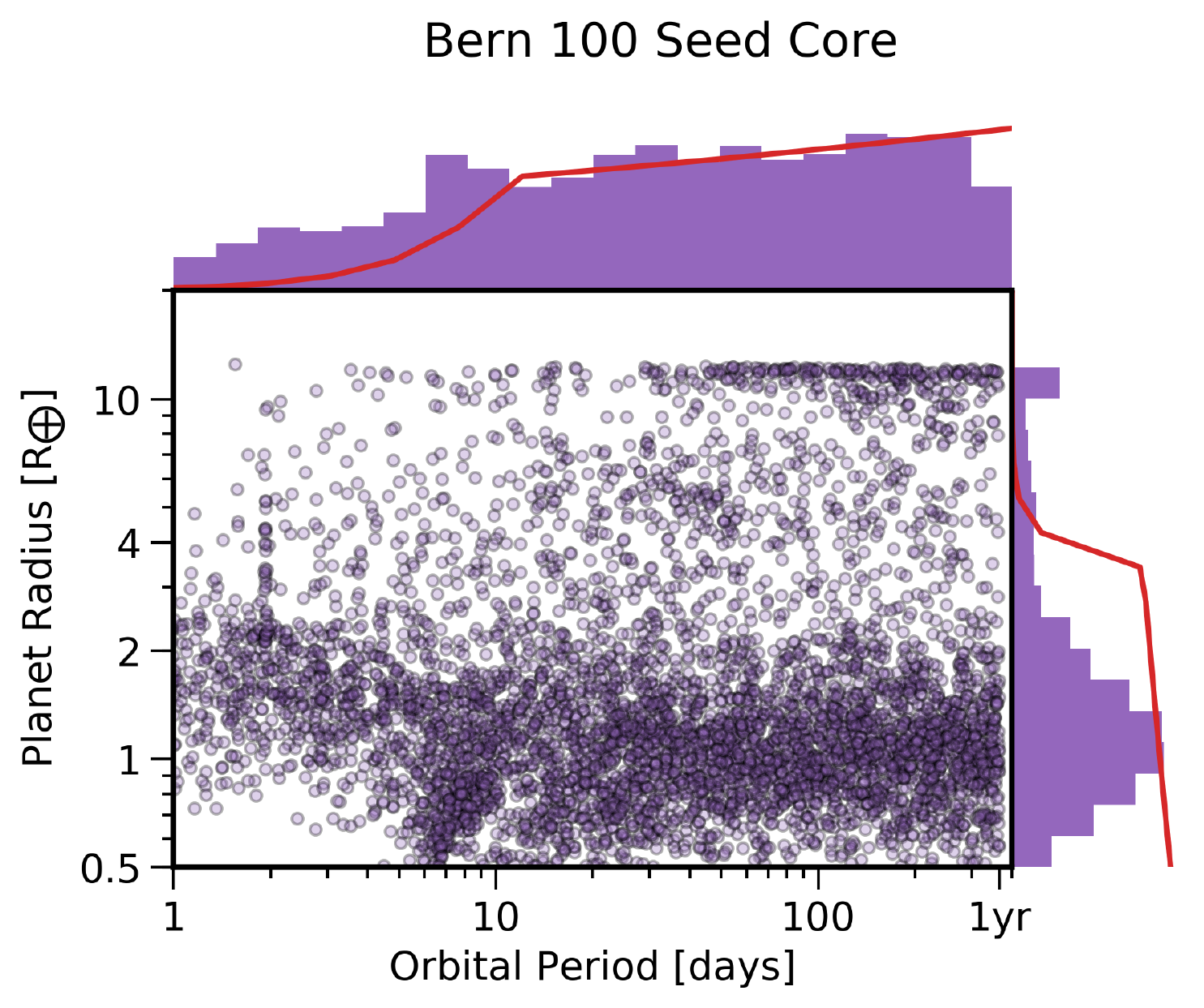}    
    \includegraphics[width=0.45\linewidth]{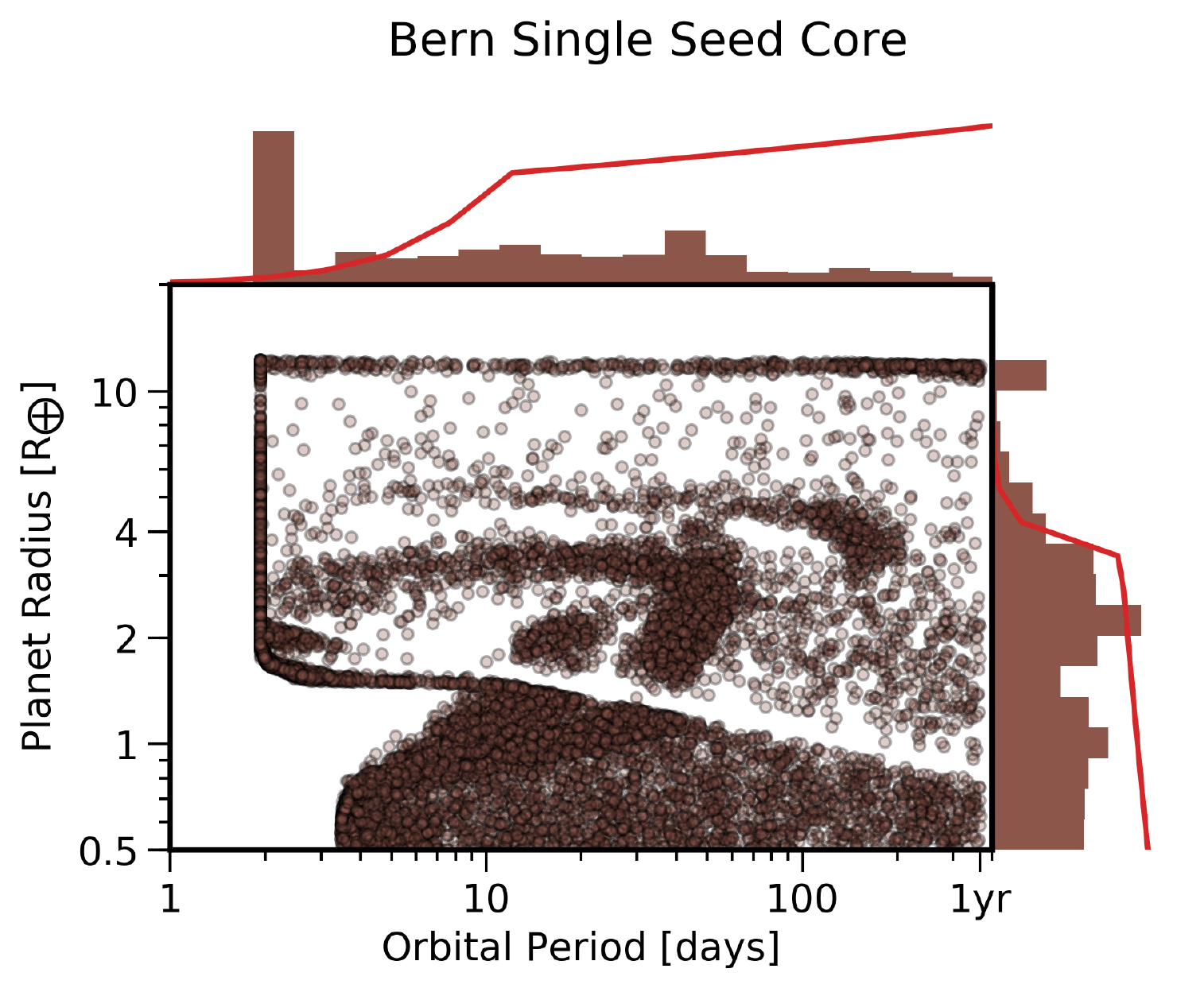}
    \setlabelscale
    \putlabel[2]{0.75}{a}
    \putlabel{0.75}{b}
    \caption{
        Planet radii and orbital periods of planets formed in the \bern population synthesis model. 
        The histograms on the top and side show the marginalized distributions compared to the \kepler parametric distributions derived in \cite{2018AJ....156...24M} (red).
          \Panel{a} shows the model with 100 seed planet cores per disk and panel \panel{b} the model with one seed planet core per disk.
        }
    \label{f:bern:PR}
\end{figure*}

\begin{figure*} 
    \centering
    \includegraphics[width=0.45\linewidth]{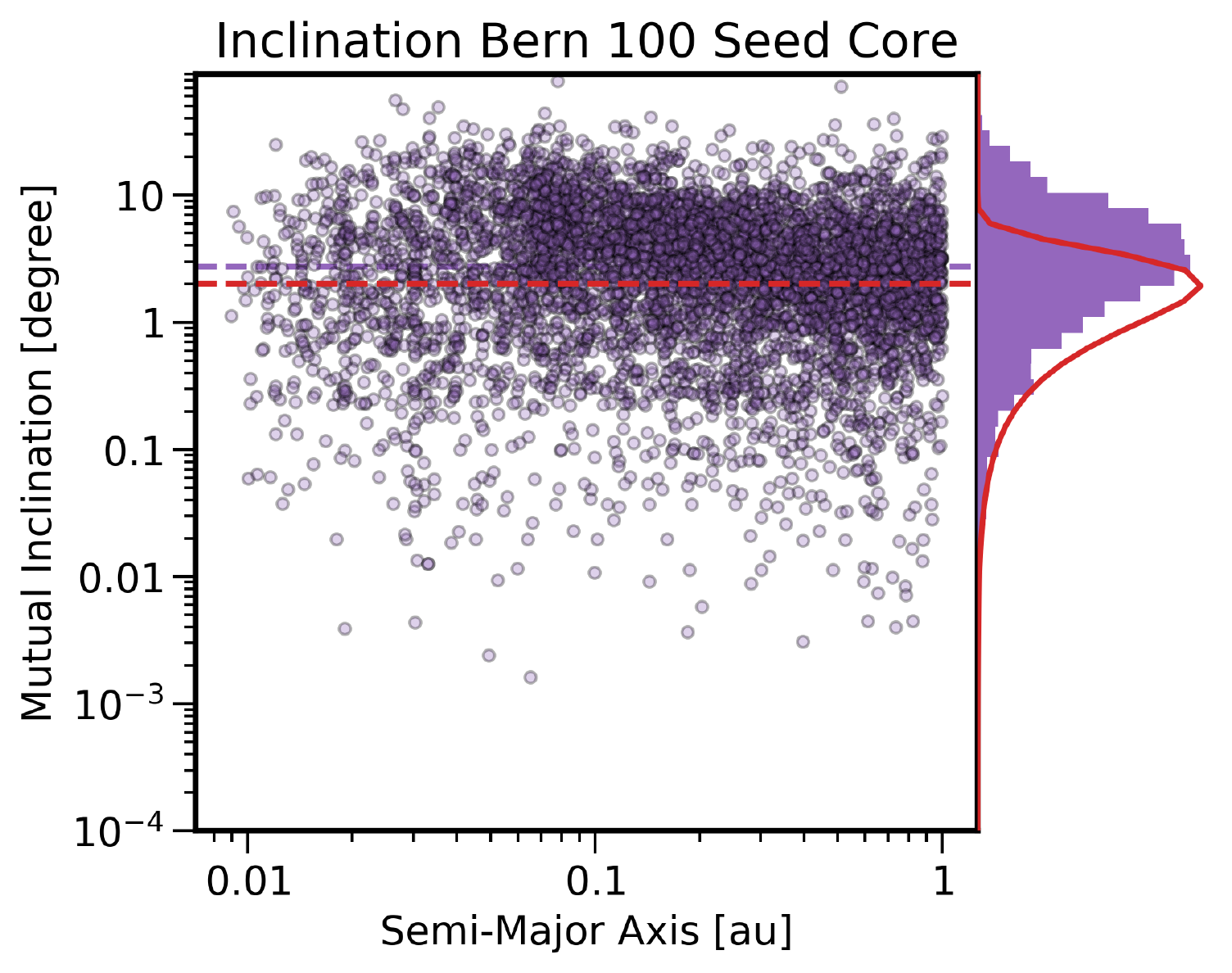}
    \includegraphics[width=0.45\linewidth]{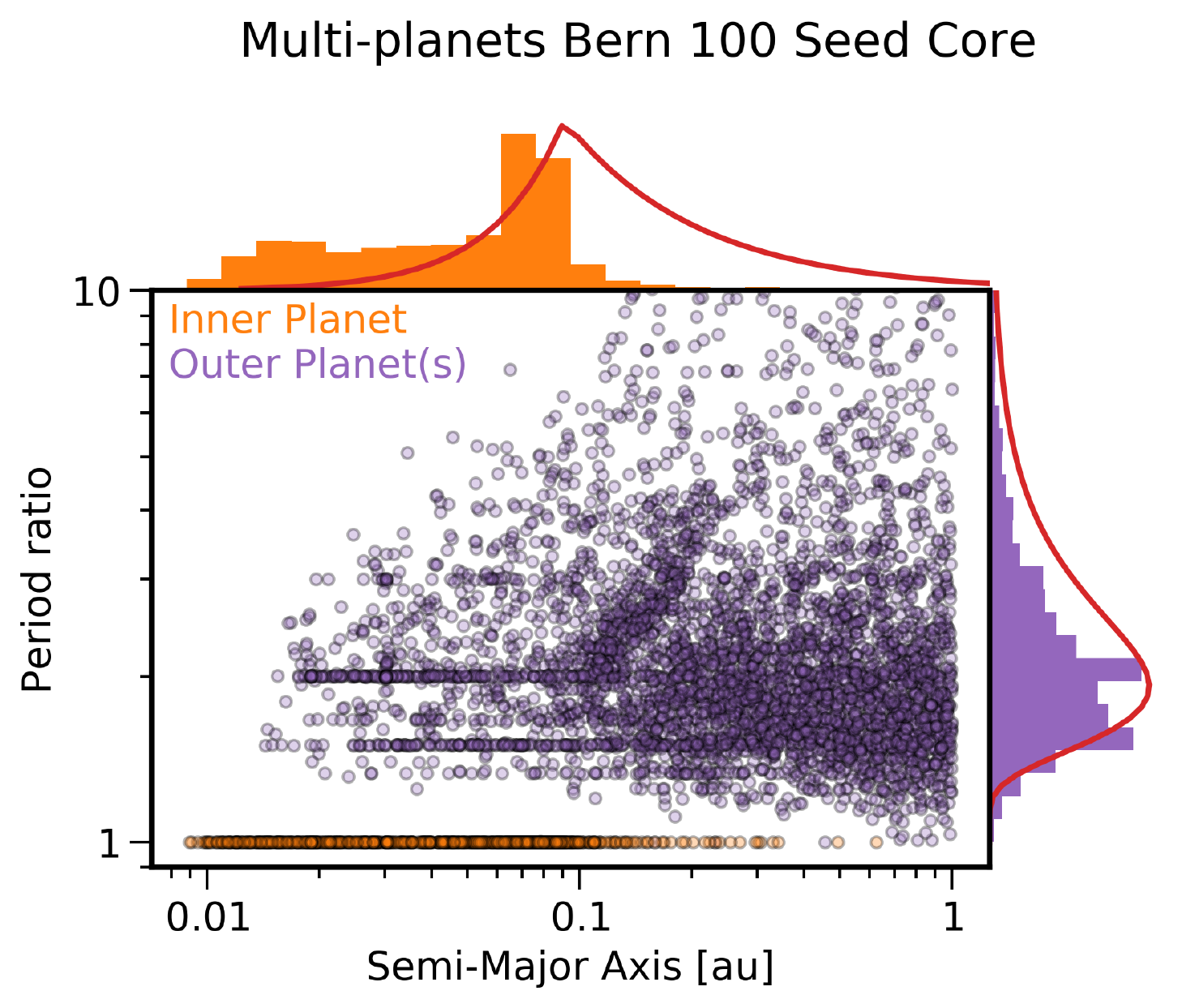}
    \setlabelscale
    \putlabel[2.1]{0.75}{a}
    \putlabel{0.75}{b}\\
    \caption{Distribution of planetary system properties of the $100$ seed-core \bern model within 1 au (purple, orange). The distributions derived from \kepler by \citetalias{2018AJ....156...24M} are shown for reference in red.
        \Panel{a} shows the mutual inclination -- defined as the planet orbital inclination with respect to an invariable plane -- versus semi-major axis. The purple dashed line shows the median mutual inclination, compared to the median mutual inclination derived from \kepler (red dashed line).
    The purple histogram on the side shows the marginalized mutual inclination distribution compared to that of \kepler (solid red line).
        \Panel{b} shows the orbital period ratios (purple) and innermost planet locations (orange).
    The period ratio of adjacent planets pairs ($\mathcal{P}=P_{k+1}/P_k$) is shown at the semi-major axis of the \textit{outer} planet in the pair ($P_{k+1}$). The purple histogram on the right shows the marginalized period ratio distribution compared to that of \kepler . 
    The semi-major axis of the innermost planet in each system is shown in orange. 
           The orange histogram on top shows the marginalized distribution of the innermost planets compared to the distribution derived from kepler (red line). 
        }
    \label{f:bern:multi}
\end{figure*}

\section{Population Synthesis Model}\label{s:bern}
In this section we briefly describe the \bern planet population synthesis models.
We use an updated version of the planet population synthesis model described in \cite{2018haex.bookE.143M} for a solar-mass host star (Emsenhuber et al. in prep).
This global model of planet formation and evolution includes the accretion of planetesimals by seed planetary cores, the gravitational interactions between planet cores using the \texttt{Mercury} integrator \citep{1999MNRAS.304..793C}, gas-driven planet migration, the accretion of gaseous envelopes from an evolving disk, and atmospheric loss through photo-evaporation and impact stripping  \citep{2005A&A...434..343A,2012A&A...540A.107F,2012A&A...547A.111M,2012A&A...547A.112M,2013A&A...549A..44F,2014ApJ...795...65J,2014prpl.conf..691B,2015IJAsB..14..201M}.
We investigate four  different sets of models, each with $1$, $20$, $50$, and $100$ seed planetary cores per disk. The models with one core per disk are mainly included for comparison with previous studies \citep[e.g.][]{2009A&A...501.1139M,2009A&A...501.1161M}, and can not be directly compared to the multi-planet systems from \kepler.

Each of the four sets consists of an ensemble of stars with a planet-forming disk. All stars are assumed to be single and of solar mass, with a composition following the observed metallicity distribution of field stars.
Every disk contains a reservoir of solids in the form of $300$ meter-sized planetesimals. The total mass in planetesimals is calculated from the disk mass and stellar metallicity.
A range of initial disk conditions is used (see \citealt{2018haex.bookE.143M}) to simulate the distribution of observed protoplanetary disk properties \citep[e.g.][]{2013ApJ...771..129A,2016ApJ...831..125P}. The distribution of disk masses spans two orders of magnitude to reflect the observed range in young protoplanetary disk masses \citep{2018ApJS..238...19T}. The total mass in planetesimals per disk ranges between $6$ and $1,500 \,M_\oplus$, with a mean of $165 \,M_\oplus$. 

Each disk is also seeded with planetary cores that can accrete those planetesimals and collide with other cores to grow into planets. 
The seed planetary cores are distributed randomly with a uniform probability in log of the distance between $0.06$ and $40$ au, with $1$, $20$, $50$, or $100$ planetary cores per disk. 
The starting masses of the seed planetary cores are $0.01 ~M_\oplus$. In all models, the seed planetary cores constitute a negligible fraction of the solid mass compared to the planetesimal disk mass.
 
N-body interactions are calculated for $10$ Myr in the 20 and 50-seed core models and for $20$ Myr for the 100 seed core model. 
This timescale is long enough to resolve system-wide instabilities after disk dispersal and form dynamically stable planetary systems within $1$ au \citep[e.g.][]{2012ApJ...751..158H}.
After that, thermodynamic evolution of the atmosphere (cooling and contraction) and atmospheric escape are tracked for $5$ Gyr. Planet radii are calculated self-consistently by solving the internal structure equations both during the formation and evolution phase, based on atmospheric accretion, loss, and cooling \citep[e.g.][]{2012A&A...547A.111M,2012A&A...547A.112M,2018ApJ...853..163J}. 
The largest changes in planetary radii occur during the first $\sim 100$ Myr, and the model does not predict significant radius evolution for the age range of stars in the \kepler field.

A thousand systems are simulated in each set of multi-seed models. 
Each set of the $20$, $50$, and $100$ seed planetary core models yields $983$, $954$, and $978$ surviving planetary systems, respectively.
Of each planetary system we use the planet radii, orbital periods, and inclinations ($\{R_p,P,\delta i\}_k$) as input to \epos. 
30,000 planets are simulated in the single seed core models that yield 29,993 planets. For the single seed model we use the set of planet radii and orbital periods ($\{R_p,P\}$) as input to \epos.

To show the typical type of output of the \bern population synthesis models, the ensemble of orbital periods and planet radii of simulated exoplanets in the 1 and 100 seed-core model are shown in \rfigure{bern:PR}. The distribution of planetary system properties of 100 seed-core model are shown in \rfigure{bern:multi}. 

The simulated ensemble of exoplanets of the 100 seed core model \sfigp{bern:PR}{a} populate the same parameter space as the \kepler exoplanets, ranging in orbital period from a day to a year and in planet size from Mars to Jupiter.
A quick comparison between the simulated exoplanets and the parametric distributions derived from \kepler in \citetalias{2018AJ....156...24M} show that the model under-predicts the occurrence of planets with sizes between 2-3 $R_\oplus$ (mini-Neptunes) and over-predicts the abundance of (rocky) planets at orbital periods less than $\sim{}10$ days.

The model with one seed planetary core per disk shows more structure in the period-radius distribution \sfigp{bern:PR}{b}. The radius distribution shows a pronounced radius valley at $\sim 1.6 R_\oplus$ as seen in the \kepler data \citep{2017AJ....154..109F,2018ApJ...853..163J}. The period-radius distribution, however, also shows many features not present in the \kepler data: In particular, the model shows a dearth of short-period earth-sized planets which are below the Type-I migration mass; 
a spike of mini-Neptunes and giants at an orbital period of 2 days corresponding to the inner edge of the computational disk at 0.03 au; 
several pile-ups (at about 20, 50 and 100 days) caused by different types of migration traps in the disk \citep{2014A&A...567A.121D};
and a spike of Jupiter-sized planets at all periods resulting from runaway gas accretion.
Otherwise, the planet radius distribution shows a similar trend as \kepler with a fairly constant occurrence of planets smaller than $3 R_\oplus$ and a sharp decrease in occurrence for larger planets.

\rfigure{bern:multi} show the typical outcome of planetary system architectures of the \bern model, here for a model with 100-seed cores. 
\sfigure{bern:multi}{a} shows the orbital inclination distribution of planets versus their semi-major axis. The inclinations show no clear trend with semi-major axis, and the marginalized inclination distribution has the same shape as the Rayleigh distribution used in the parametric solution of \citetalias{2018AJ....156...24M}. 
\sfigure{bern:multi}{b} shows the period ratios, \dpr, of adjacent planet pairs as a function of semi-major axis. It also shows the location of the innermost planet in each planetary system. The simulated period ratio distribution becomes wider at larger semi-major axes, but there is no clear sign of a covariance between period ratio and star-planet separation. There are concentrations of planets at the 2:1 and 3:2 orbital resonances.

The distributions of inclinations, period ratios, and innermost planet locations of the \bern model follow wide but peaked distributions that are similar in shape to the distributions observed or inferred from \kepler. However, there are also a number of key differences between the observations and the simulated systems in the 20, 50, and 100 core models that we will discuss in depth in the next sections.

\begin{figure*}
    \includegraphics[width=0.45\linewidth]{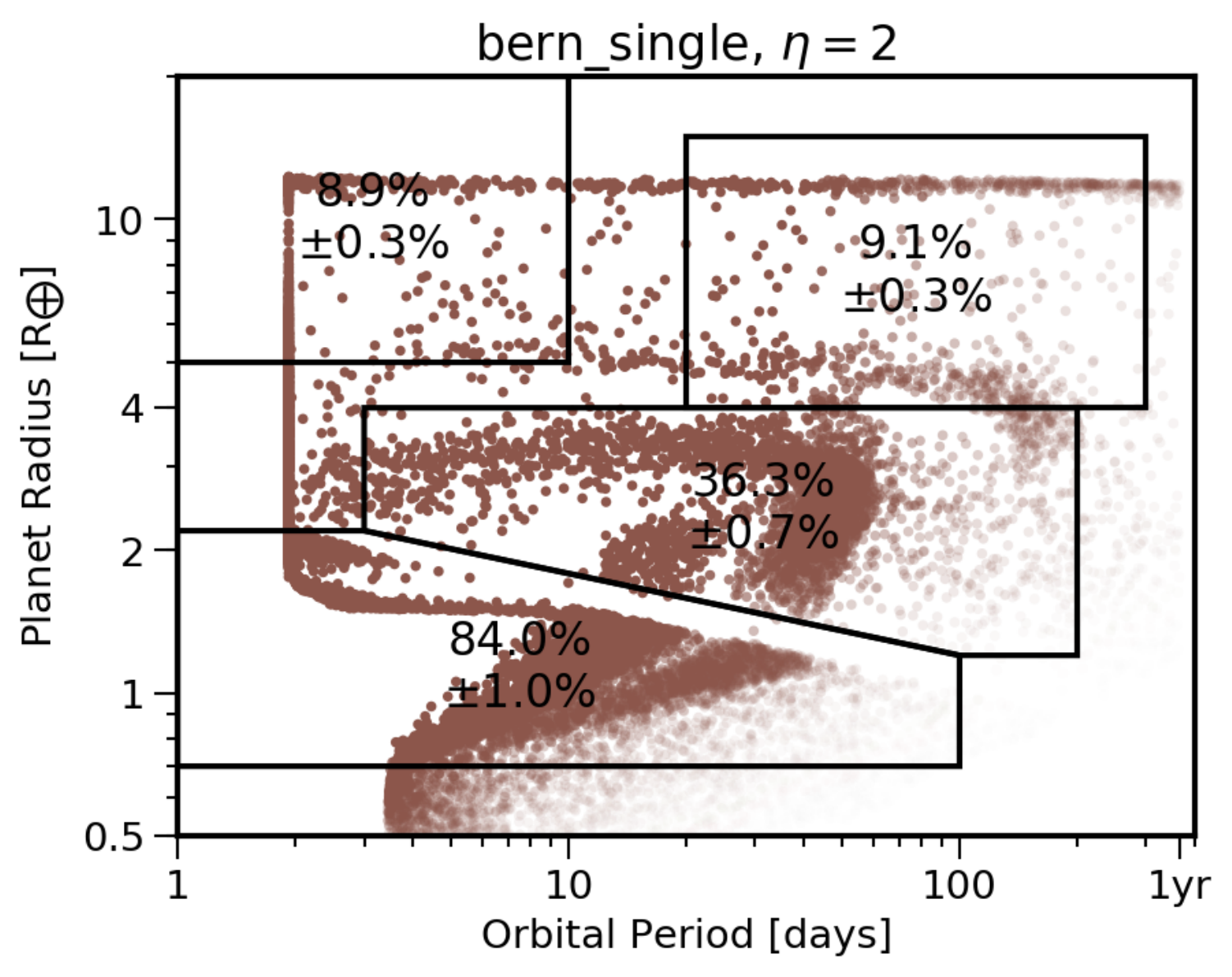}
    \includegraphics[width=0.45\linewidth]{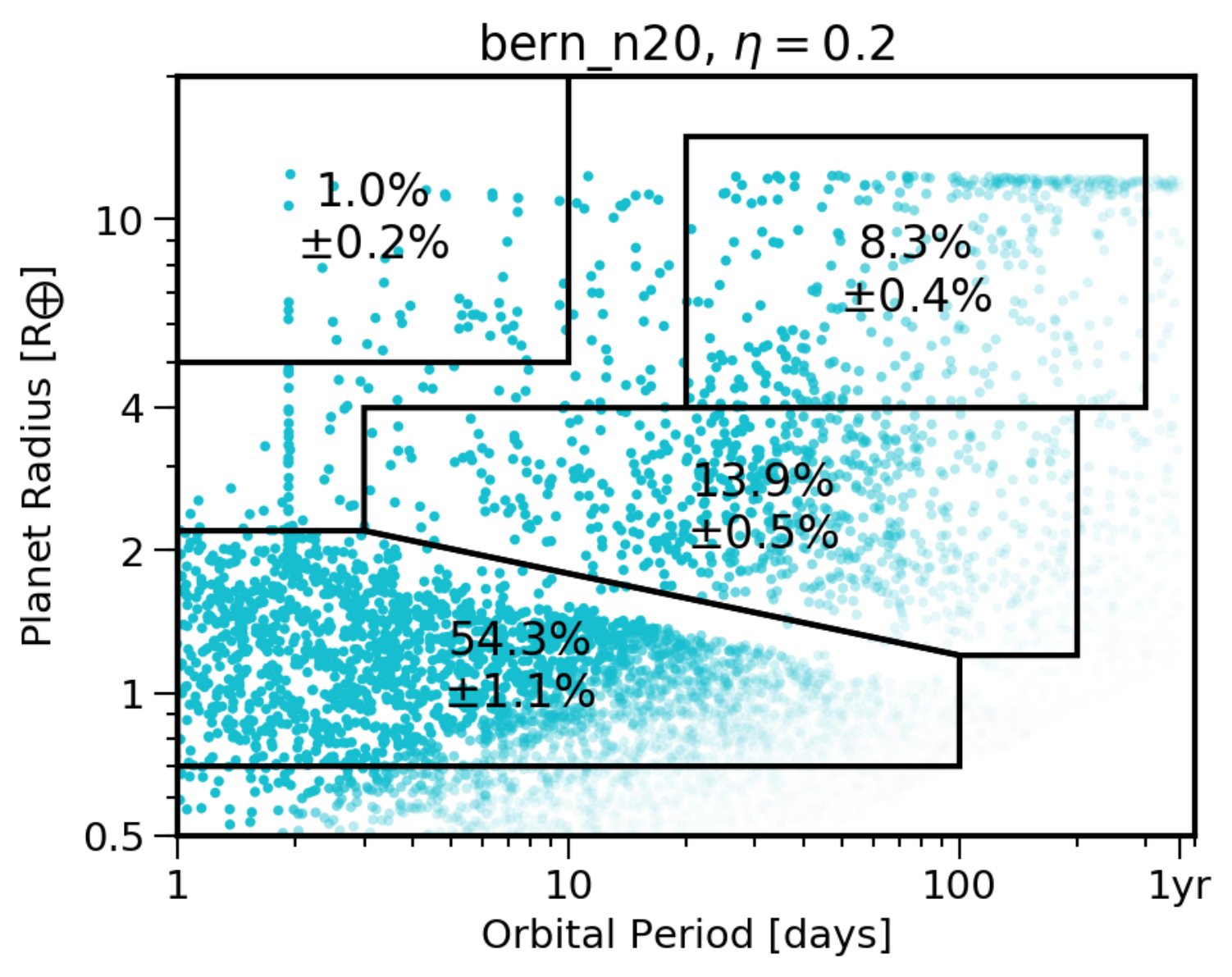}
    \setlabelscale
    \putlabel[2]{0.75}{a}
    \putlabel{0.75}{b}\\
    \includegraphics[width=0.45\linewidth]{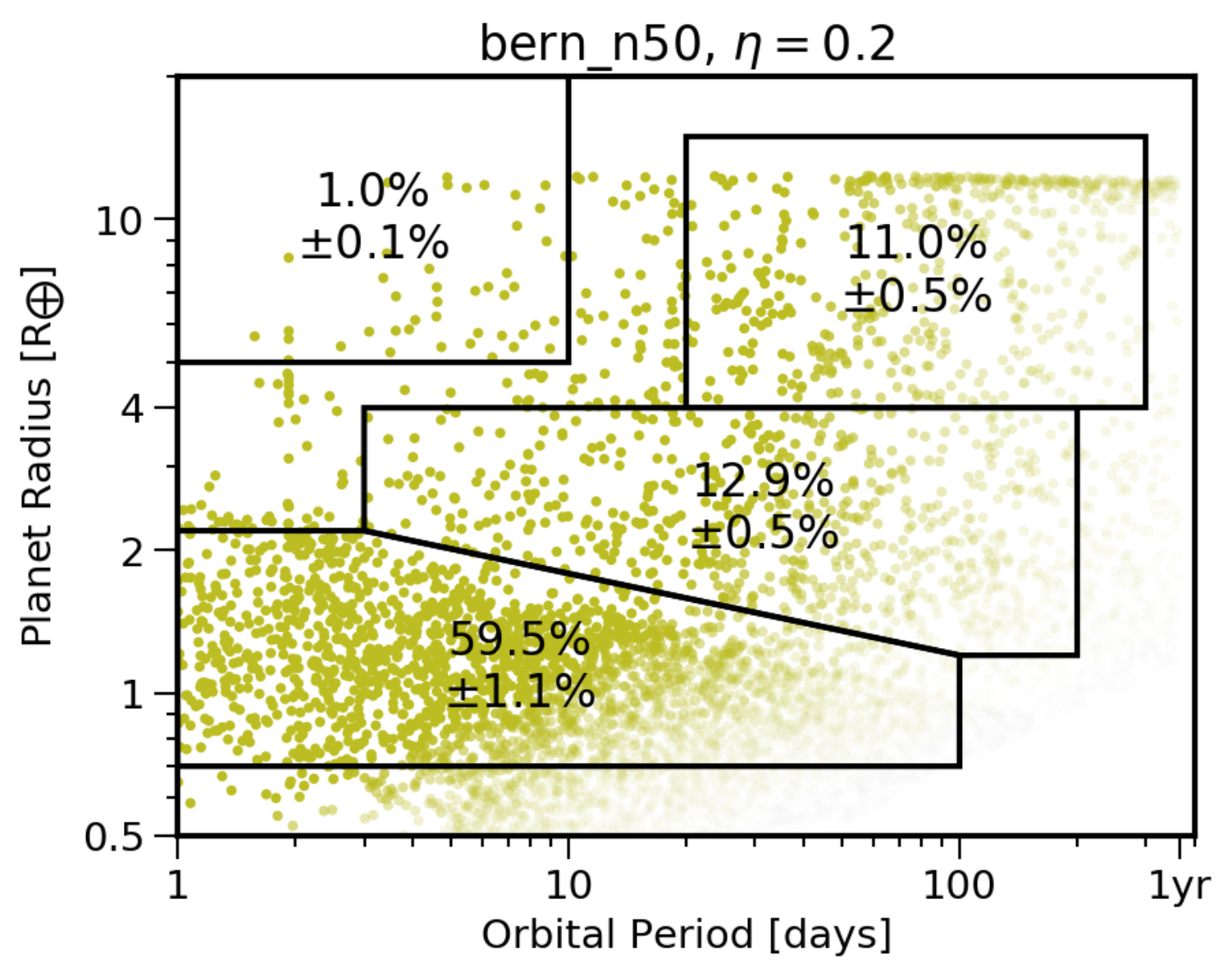}
    \includegraphics[width=0.45\linewidth]{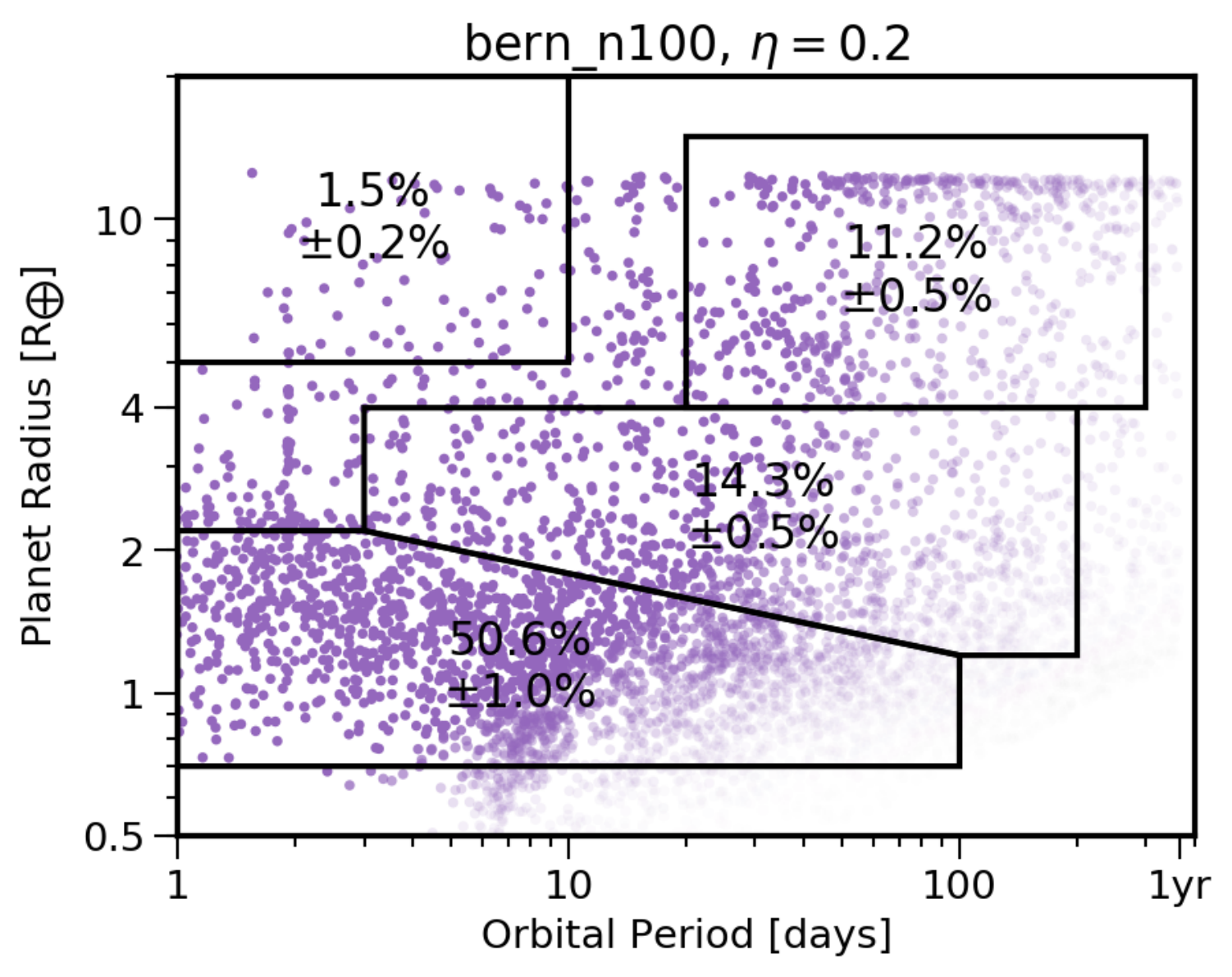}
    \setlabelscale
    \putlabel[2]{0.75}{c}
    \putlabel{0.75}{d}\\
    \includegraphics[width=0.45\linewidth]{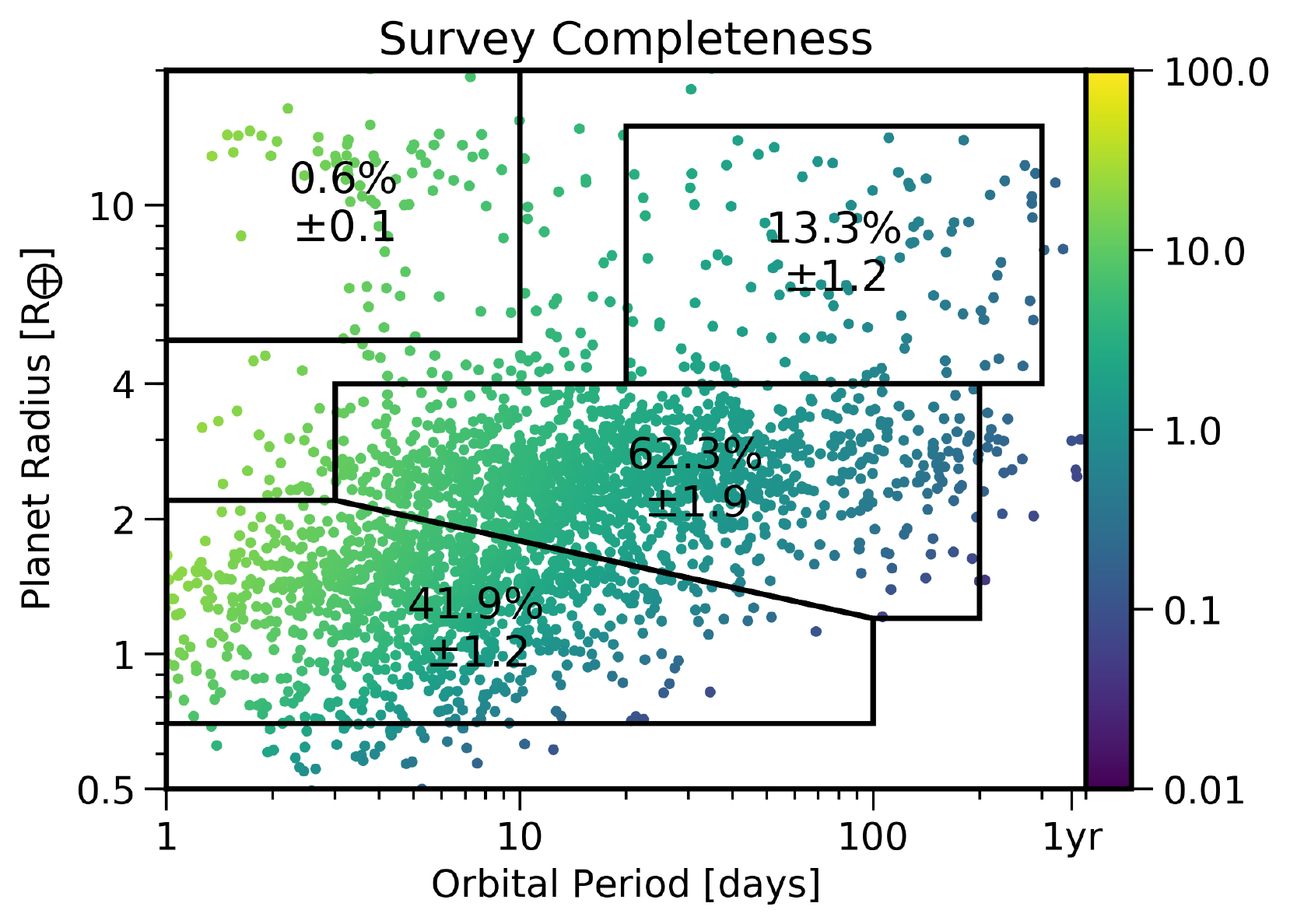}
    \includegraphics[width=0.45\linewidth]{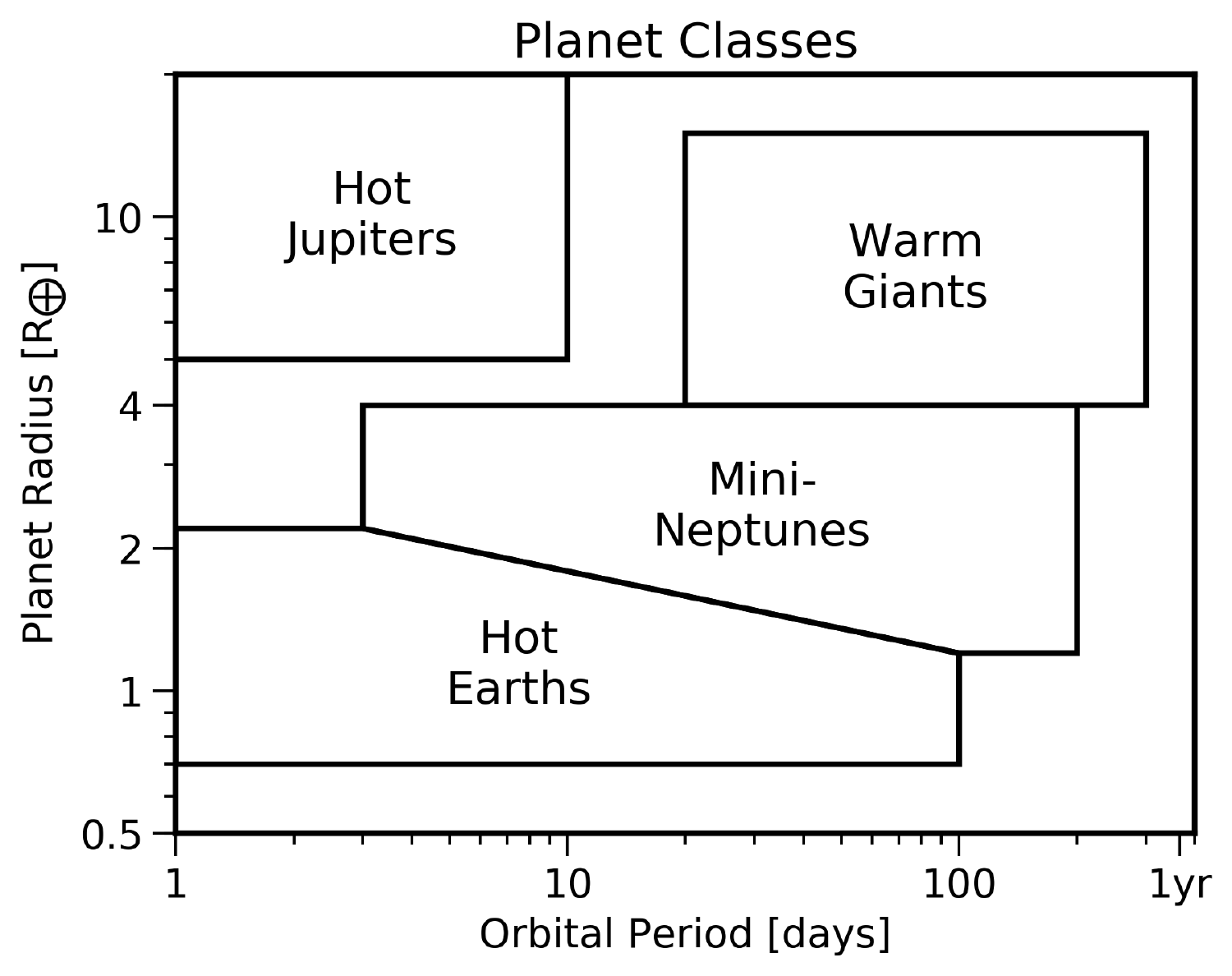}
    \setlabelscale
    \putlabel[2]{0.75}{e}
    \putlabel{0.75}{f}\\
        \caption{Planet occurrence rates of the planet population synthesis model \panelp{abcd} compared to those of Kepler \panelp{e} for different planet categories defined in \panelp{f}. 
        Bins span equal areas in logarithmic area units ($d \log R ~d \log P$) and roughly correspond to hot Jupiters, warm giants, super-earths, and mini-Neptunes.
        \textbf{The models yield occurrence rates similar to observations for Warm Giants, Hot Jupiters, and Hot Earths, but systematically underestimate the occurrence of mini-Neptunes compared to the observed rates.}
    }
    \label{f:occ}
\end{figure*}

\section{Planet Occurrence Rates}\label{s:occ}
The next step is to compare the model's planet occurrence rates for different classes of planets to the \kepler data. The occurrence rate in a certain period-radius area is defined as the total number of simulated planets in that area across all simulations divided by the number of simulated stars, multiplied by a factor $\eta$, the fraction of stars that have planetary systems akin to those modeled here. The factor $\eta=0.2$ is chosen such that the total number of detectable planets is roughly equal to that of \kepler. For the simulations with one seed planetary core per disk, we set $\eta=2$ because stars have, on average, more than one planet. 

In \rfigure{occ} we show the simulated populations as they would appear when observed in the \kepler survey. 
\sfigure{occ}{e} shows the exoplanet candidate list color-coded by the survey completeness. The planet occurrence rates are then calculated as the sum of the inverse completeness of each planet in the area divided by the number of stars in the survey, $n_\star= 122,682$. 

\begin{figure}
    \centering
    \includegraphics[width=\linewidth]{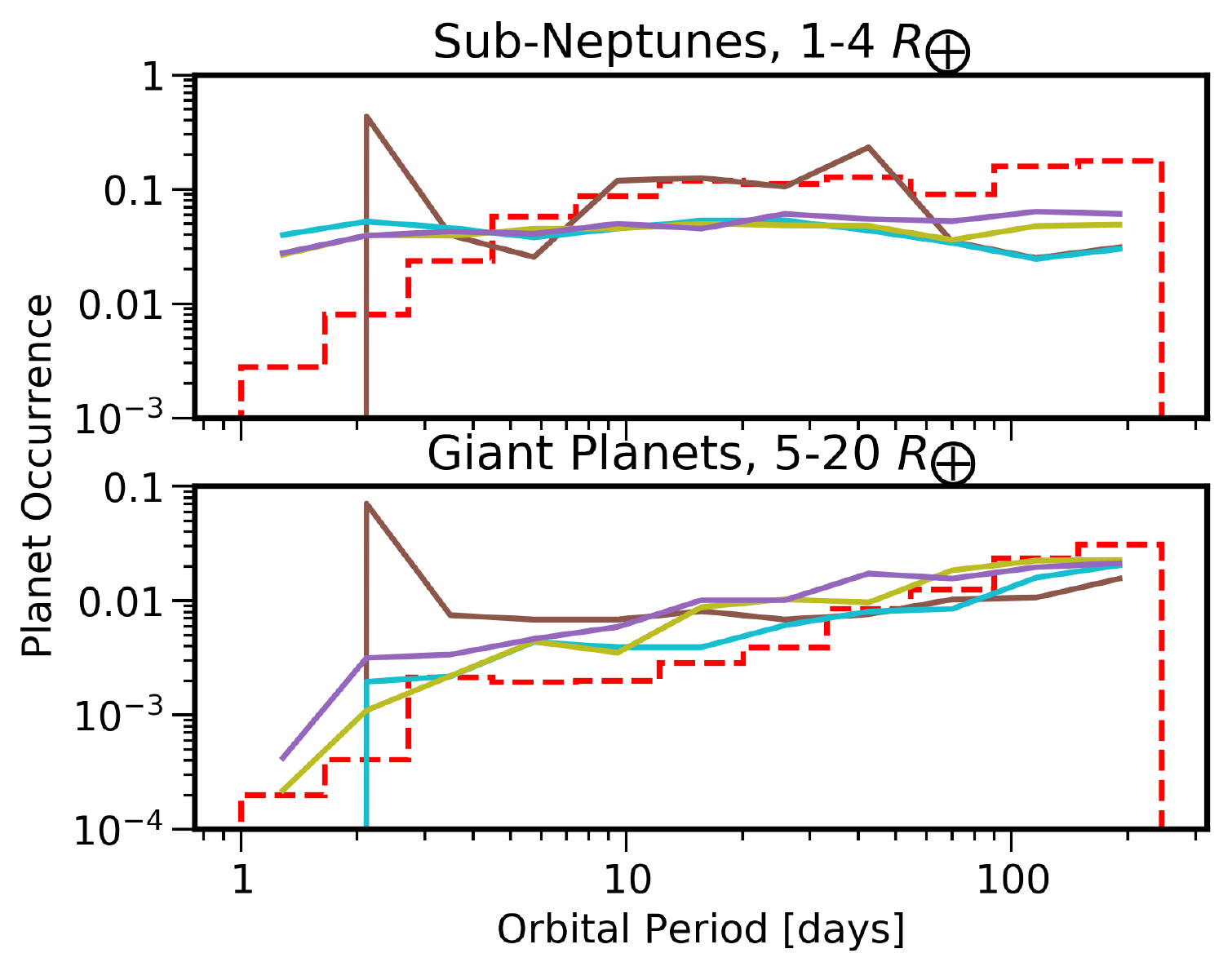}    
    \setlabelscalesingle
    \putlabel{0.75}{a}\\
    \includegraphics[width=\linewidth]{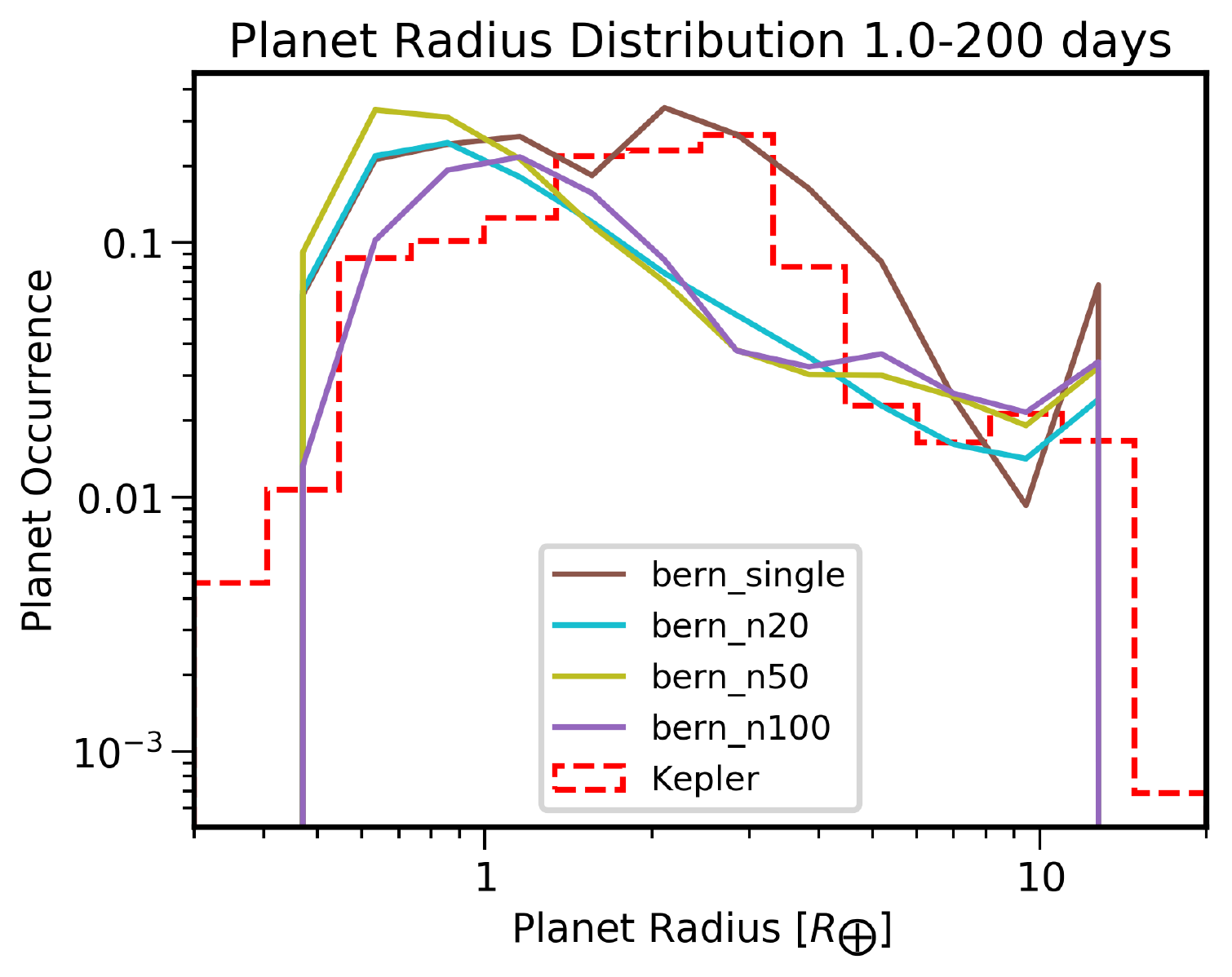}
    \putlabel{0.75}{b}
    \caption{
    Planet occurrence rates as function of planet radius \panelp{a} and orbital period for giant planets and sub-Neptunes \panelp{b}. 
    }
    \label{f:occrates}
\end{figure}

Here, we define occurrence rates of different classes of planets that roughly compare to the observed clusters in the \kepler data, \sfig{occ}{e}:
\begin{itemize}
\item Hot Jupiters: $R=5-20~R_\oplus$, $P=1-10$ days.
\item Warm Giants: $R=4-14~R_\oplus$, $P=20-300$ days.
\item Hot Earths: $R=0.7-2.2~R_\oplus$, $P=3-200$ days, and below the radius valley.
\item Mini-Neptunes: $R=1.2-4.0~R_\oplus$, $P=3-200$ days, and above the radius valley.
\end{itemize}

The period and radius ranges are chosen to span roughly equal areas in log space, with $d \ln P d \ln R_p\approx 3.1$. Hot and warm giant planets are separated by the ``period valley'' at $\sim 10$ days \citep{2016A&A...587A..64S}. Giant planets are separated from sub-Neptunes at $4~R_\oplus$. Sub-Neptunes are split into Hot Earths and mini-Neptunes at the evaporation valley \citep{2017AJ....154..109F}. Bins are mostly bounded by detection limits at the largest periods and the smallest radii where planets have been detected. 
Most notably these bins exclude the habitable zone, though we will make inferences based on the simulated planet populations in Section \ref{s:eta}. 

The \kepler planet occurrence rates indicate that mini-Neptunes are the most populous class of planets at $62.3\%$, followed by Hot Earths ($42\%$), Warm Giants ($13\%$), and last and least Hot Jupiters ($0.6\%$). The confidence intervals are based on poisson statistics only, and do not account for systematic uncertainties in the planet occurrence rate calculation.

Planet occurrence rates for the 20, 50, and 100 seed models \sfigp{occ}{bcd} are calculated assuming 1 in 5 stars have a simulated planetary system ($\eta=0.2$). Despite the different number of seed cores, the overall planet occurrence rates as function of orbital period and planet radius are very similar. This indicates that the number of planets formed in each simulation is not dependent on the amount of seed cores but on other properties that do no vary between the different sets of simulation, such as the mass/amount of planetesimals that can be accreted. 
All three multi-seed models roughly produce the same planet occurrence rates for Hot Jupiters (1-1.5\%), Warm Giants (8-11\%), mini-Neptunes (13-14\%), and Hot Earths (51-60\%). The simulated rates approach the derived planet occurrence rates from \kepler within a factor of 2, with the exception of mini-Neptunes which the models under-predicts by a factor 4-5.

The single seed model predicts a similar occurrence rate of Warm Giants as the multi-seed models and as the \kepler data when $\eta=2$ planets each are assigned to the simulated stars. The occurrence rate of mini-Neptunes ($36\%$) is thrice that of the multi-seed models but still twice as low as \kepler. However, the single seed model over-predicts the number of hot Jupiters by more than a factor 10. The model also over-predicts the number of Hot Earths by a factor $2$ at $84\%$. It should be noted that the clustering of rocky planets in period-radius space is very different from the observed smooth distribution of \kepler planets. The occurrence rate of hot Earths is therefore very sensitive to the locations of the bin boundaries in \rfigure{occ}.

We also show the planet occurrence rates as a function of planet radius and orbital period in \rfigure{occrates}.
The giant planet occurrence rates in the multi seed-core models \sfigp{occrates}{a} show the same increasing trend with orbital period as the \kepler planet occurrence rates (see also \citealt{2019ApJ...874...81F}). The single seed models show a pronounced peak for hot-Jupiters at $P\approx 3$ days, corresponding to a very strong trap for migrating planets at the disk inner edge, that is not seen in the \kepler data. Sub-Neptunes show a flat trend with orbital period, and do not capture the observed decrease in occurrence interior to 10 days \citep{2011ApJ...742...38Y,2015ApJ...798..112M} which is likely due to a migration trap at the inner edge of the protoplanetary disk \citep{2017ApJ...842...40L,2019MNRAS.tmp..963C}. This indicates that the planet trap in the model is located to close to the star and possibly that too many planets are pushed further in when captured in resonances with other migrating planets, as no decrease at the shortest periods is seen either. 

The planet radius distribution shows the same overall trend as the \kepler distribution, with sub-Neptunes occurring roughly ten times as frequent as giant planets \sfigp{occrates}{b}. The radius distribution among sub-Neptunes, however, has a distinctly different shape than the observed distribution. The multi-seed models predict a much higher fraction of earth-sized planets compared to mini-Neptunes ($2-3 R_\oplus$) than observed. The single-core model produces a larger amount of mini-Neptunes that fits better to the observed distribution. But it also shows a pronounced peak of giant planets at $\sim 10 \,R_\oplus$ that is not seen in the \kepler data. The observed break in the radius distribution at $\sim 3 R_\oplus$ is reproduced by the single seed model, while the multi-seed models show a more gradual decrease in occurrence in the $1-4 R_\oplus$ range.

The dearth of mini-Neptunes in the simulated multi-seed models highlights a key component of the new \bern planet population synthesis models. As proto-planets dynamically interact and compete for material, their growth process diverges from that of an isolated planet. While multi-planet interactions result in fewer planets undergoing type-II migration and becoming hot Jupiters, it also suppresses the growth of mini-Neptunes to a level that appears inconsistent with the observed abundances. The ratio of mini-Neptunes to rocky planets is larger in the single-seed models compared to the multi-seed models \sfigp{occrates}{b}.
Increasing the gas/dust mass of the disk may not be a solution, as this may trigger more planets to reach runaway growth and form giant planets, potentially overshooting the amount of warm giants and/or hot Jupiters. A more efficient formation pathway for accreting hydrogen atmospheres without ballooning into giant planets appears necessary. This was also pointed out by \cite{2018ApJ...869L..34S} for planets at larger separations probed by microlensing surveys.

The evaporation valley also appears to get weaker when including more seed planetary cores, either through atmospheric stripping or through dynamical scattering, providing a potential diagnostic for calibrating the amount of dynamical interaction.

%
%

\section{Planetary System Architectures}\label{s:multi}
Here we inspect the architectures of multi-planet systems formed in the 20, 50, and 100-seed core models and evaluate the ability of the \bern model to reproduce planetary system architectures observed with \kepler. We make two types of comparisons between the model and the data. One method is to put the planetary system properties in context of the debiased parametric distributions derived from \kepler. The other method is to simulate a survey using the planetary systems from the \bern model and compare its observable properties directly to the \kepler data.
Both methods have their pros and cons, and we will therefore present them side by side. 

The parametric distributions are more intuitive as they directly tell you what the distribution of planet formation model outcomes should be. The major downsides of the parametric models of the exoplanet population, however, is that they make a number of simplifying assumptions that may not be justified by real planetary systems or by the formation model. In particular, the parametric distributions derived in \citetalias{2018AJ....156...24M} do not allow for correlations between different parameters. The forward modeling approach is also better for recognizing and identifying clusters and valleys or discontinuities in distributions that are not specifically included in the parametric distributions, such as orbital resonances.

We focus on three properties of the model planetary systems and their associated observable summary statistic. 
\begin{enumerate}
\item The inclination distribution of planets within a planetary system (e.g. their mutual inclinations) whose observable is the frequency of multi-planet systems.
\item The period ratio of adjacent planets in a system, whose (biased) distribution can be observed.
\item The location of the innermost planet in a planetary system, whose (biased) distribution can also be observed.
\end{enumerate}
An example of these summary statistics for the 100 seed-core model and how they relate to the instrinsic model properties is shown in the appendix. All summary statistics are calculated for orbital periods between $1$--$400$ days and planet radii between $0.5$--$20 ~R_\oplus$.
These summary statistics mainly trace the dynamical interactions between forming protoplanets and the trapping of migrating planets at the inner edge of the protoplanetary disk. 

\begin{figure*} 
    \centering
    \includegraphics[width=0.4\linewidth]{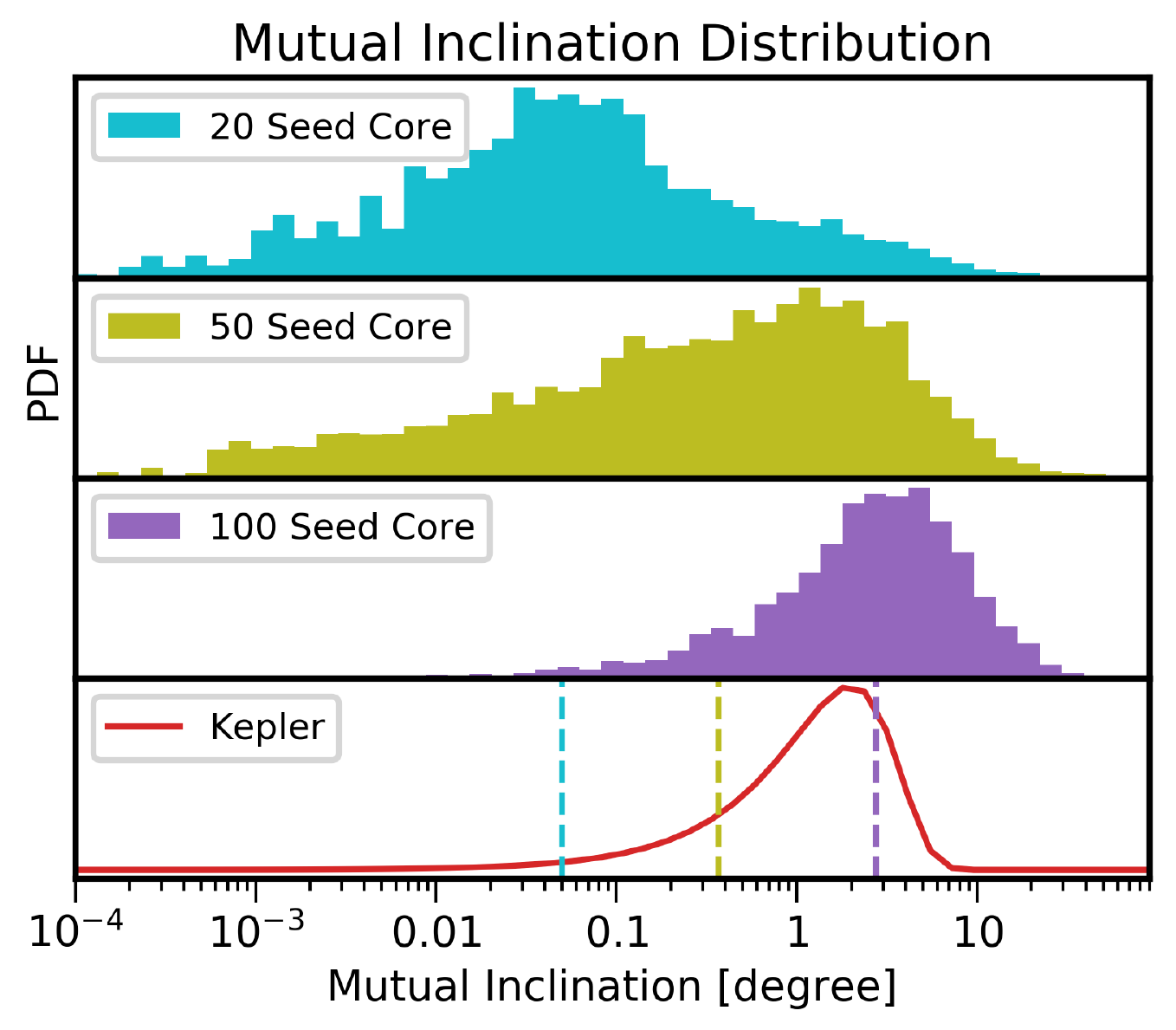}
    \includegraphics[width=0.45\linewidth]{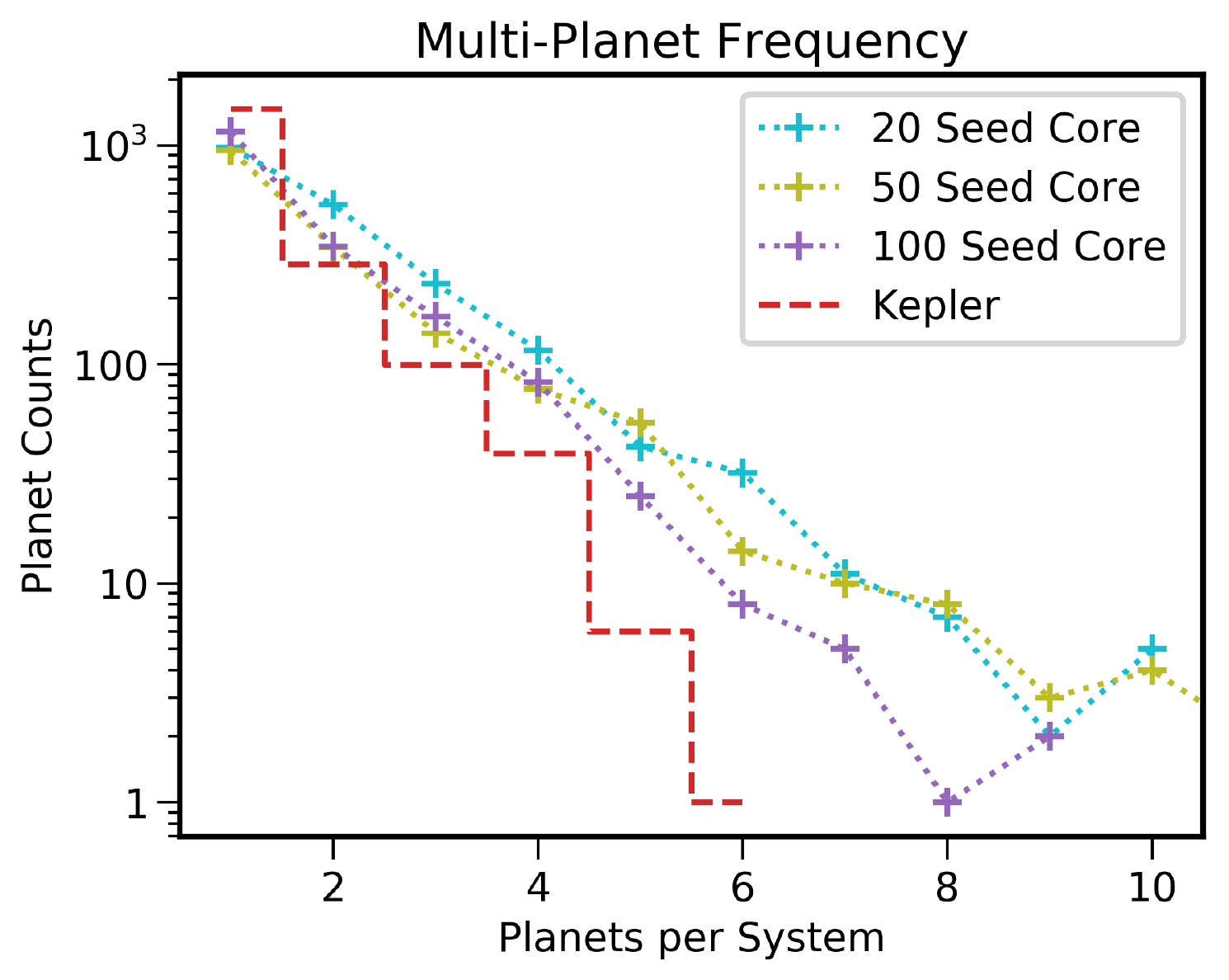}
    \setlabelscale
    \putlabel[2]{0.75}{a}
    \putlabel{0.75}{b}
     \caption{Mutual inclination distribution and observable multi-planet frequencies.
     \Panel{a} shows the mutual inclination distributions for the 20, 50, and 100 seed-core models, here defined with respect to the invariable plane of the system. The median mutual inclination for each model is indicated in the bottom panel with a dashed vertical line. The best-fit Rayleigh distributions from \kepler derived by \citetalias{2018AJ....156...24M} is shown for comparison (solid red line).
     \Panel{b} shows the frequency of multi-planet systems with $k$ planets in a simulated survey of the \bern models. The multi-planet frequency from \kepler in the same radius and period range is shown with the dashed red line.
     \bf Models with more seed cores per disk have higher mutual inclinations, but all models still over-predict the frequency of systems with multiple planets compared to the observed \kepler systems.
    }
    \label{f:all_inc}
\end{figure*}

\subsection{Mutual inclinations}
The mutual inclinations between planets in multi-planet systems trace the amount of dynamical interactions during the late stages of planet formation. Orbital damping by gas keeps the planets' mutual inclinations low, while gravitational interactions between (proto)planets tend to increase their mutual inclinations. The \kepler systems, with a typical mutual inclination of a few degrees \citep[e.g.][]{2014ApJ...790..146F}, show evidence of both damping and excitation \citep[e.g.][]{2016ApJ...822...54D}.

The inclination distributions of the 20, 50, and 100 seed core models are broad but with a clear peak \sfigp{all_inc}{a}. The distribution narrows when a larger number of cores are included in the model, and hence more dynamical interactions take place. The median inclination also increases from $\delta i=0.05\degr$ in the 20 core model to $\delta i=0.4\degr$ in the 50 core model to $\delta i=3\degr$ in the 100 core-model. The inclination distribution of the 100-core model has the same shape as the Rayleigh distribution used in the parametric solution, while the distributions with fewer cores are wider and have a tail towards very low inclinations of $10^{-4}~\degr$. 
These extremely low inclinations indicate that some protoplanets may not experience gravitational interactions if the number of seed cores in the simulation is low.

The median inclination of the 100 core model is close to that derived from \kepler, while the inclinations of the 20 and 50 core models are significantly lower. This is reflected in the simulated observations of the multi-seed models \sfigp{all_inc}{b} that predict fewer multi-planet systems when more cores are included. The simulated observations of the multi-seed models, however, all predict a larger number of systems with two or more planets per star compared to what is observed with \kepler.
The relative frequencies of 2/3 and 3/4 planets-per-system in the simulated observations match the observed relative frequencies. 

The predicted ratio of observed single to double planet systems, however, is significantly smaller than observed. This effect is commonly referred to as the \kepler dichotomy as it may hint at an additional population of planet(ary system)s \citep{2012ApJ...758...39J,2013ApJ...775...53H,2016ApJ...816...66B,2016ApJ...832...34M}. However, the dichotomy may also be an artifact of the planet detection pipeline \citep{2019MNRAS.483.4479Z} or a complex feature in the architectures of planetary systems \citep{2018ApJ...860..101Z,2019MNRAS.tmp.2033S} -- though we do not see such a feature arise in the \bern models. We can account for this dichotomy following the procedure in \cite{2018AJ....156...24M}, i.e. by assigning isotropic mutual inclinations to a fraction of simulated planetary systems. The ratio of 1/2 planets-per-system, would lead to a better match to the data if $f_\text{iso}\approx 40\%$ of the systems are assumed to have high mutual inclinations. This value is consistent with the values derived in \cite{2018AJ....156...24M,2019MNRAS.490.4575H}, but significantly larger than the $5-19\%$ effect reported by \cite{2019MNRAS.483.4479Z}. 

The frequencies of systems with 5 or more planets would still be over-predicted in the simulated observations after correcting for the dichotomy. Most likely, the large number of simulated planets close to the star (see also \S \ref{s:inner}) increase the number of observable systems with large numbers of planets per system. We will address this in a future iteration of the \bern model.

\begin{figure*} 
    \centering
    \includegraphics[width=0.4\linewidth]{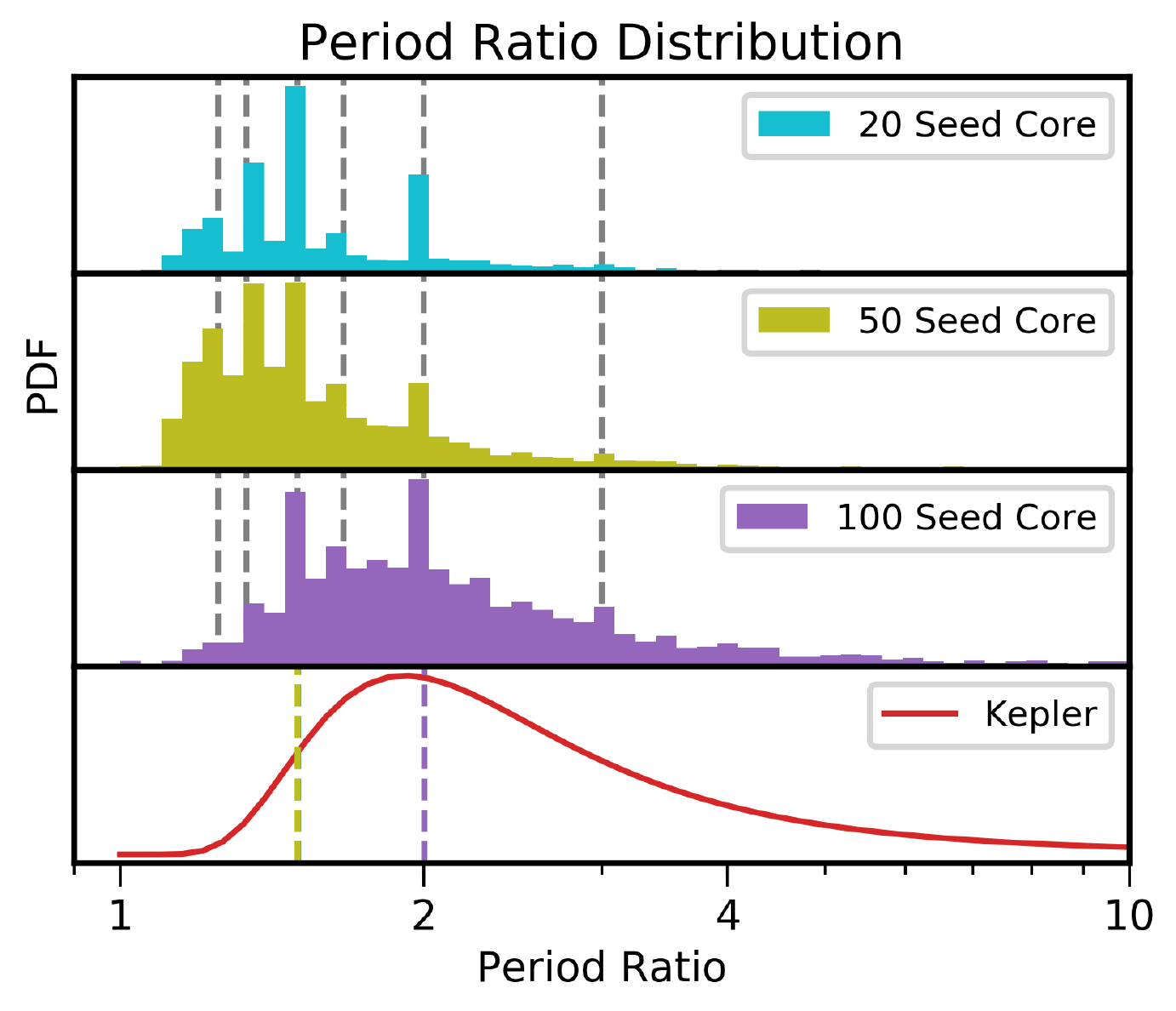}
    \includegraphics[width=0.45\linewidth]{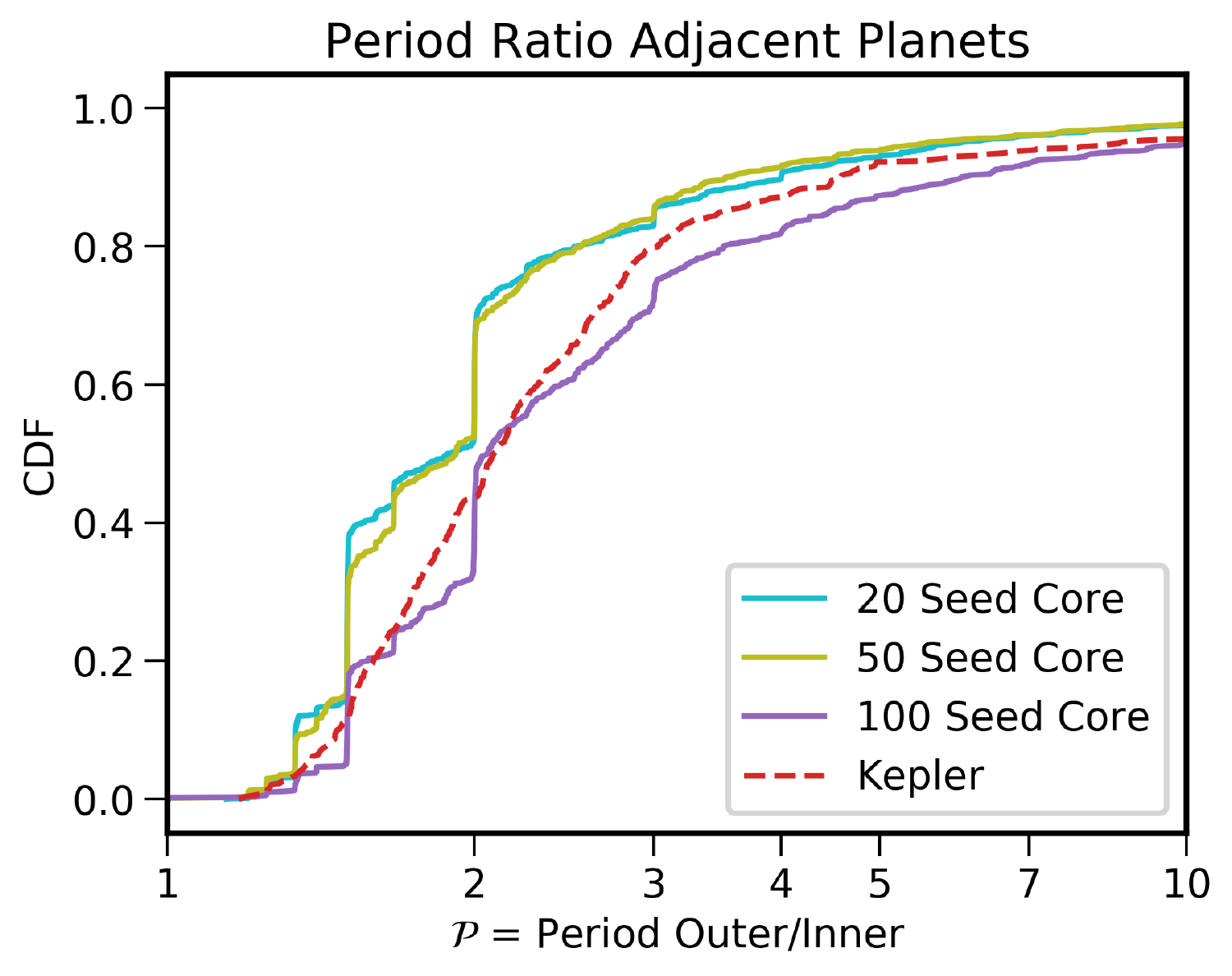}
    \setlabelscale
    \putlabel[2]{0.75}{a}
    \putlabel{0.75}{b}
     \caption{Orbital period ratio distribution of the \bern models compared to those of \kepler. 
    \Panel{a} shows the intrinsic period ratio distribution predicted by the model compared to the parametric distribution derived from \kepler.
    The dashed gray lines indicate orbital resonances, the dashed colored lines indicate the mean orbital period ratio of the models.
    \Panel{b} shows cumulative distribution of period ratios in a simulated survey compared to the distribution observed with \kepler.
    \textbf{Models with a larger number of seed cores provide a better match to the observed period ratio distribution but over-predict the number of planets in orbital resonances.}
    }
    \label{f:all_dP}
\end{figure*}

\subsection{Period Ratios}
The period ratio between adjacent planets is another tracer of the amount of dynamical excitation and orbital damping during planet assembly. More gravitational interactions between protoplanets lead to wider spacings and fewer orbital resonances, while more damping by the gas leads to smaller spacings and more orbital resonances. 

The period ratios of the multi-seed models follow a broad distribution with pronounced peaks at the location of orbital resonances \rfigp{all_dP}. The period ratio distribution of the 100 seed-core model is most similar in shape to the \kepler distribution, with a median at $\mathcal{P} \sim 2$ and a long tail towards large period ratios \sfigp{all_dP}{a}. The 20 and 50-core models have a period ratio distribution that peaks at shorter period ratios than \kepler, both with a median at $\dpr \sim 1.5$. Orbital resonances, indicated with the dashed gray lines, appear more prominent when fewer planetary seed cores are considered, hinting at fewer gravitational interactions between forming planets in those cases.

The overall shape of the period ratio distribution in a simulated survey of the 100 seed-core model, shown as a cumulative distribution in \sfigp{all_dP}{b}, is similar to the observed one. While the 20 and 50 core models have on average smaller period ratios than observed, the 100-core model is a close match to the observed distribution, indicating the role that increased dynamical interactions play in making the synthetic simulated populations more closely match the distribution observed with \kepler. 
This observational comparison clearly favors the 100 seed-core model. However, the period ratio distribution shows more structure around orbital resonances than the \kepler data, implying that more dynamical interactions between planets are needed to push more planets out of resonances and smooth out these structures.

 \begin{figure*} 
    \centering
    \includegraphics[width=0.4\linewidth]{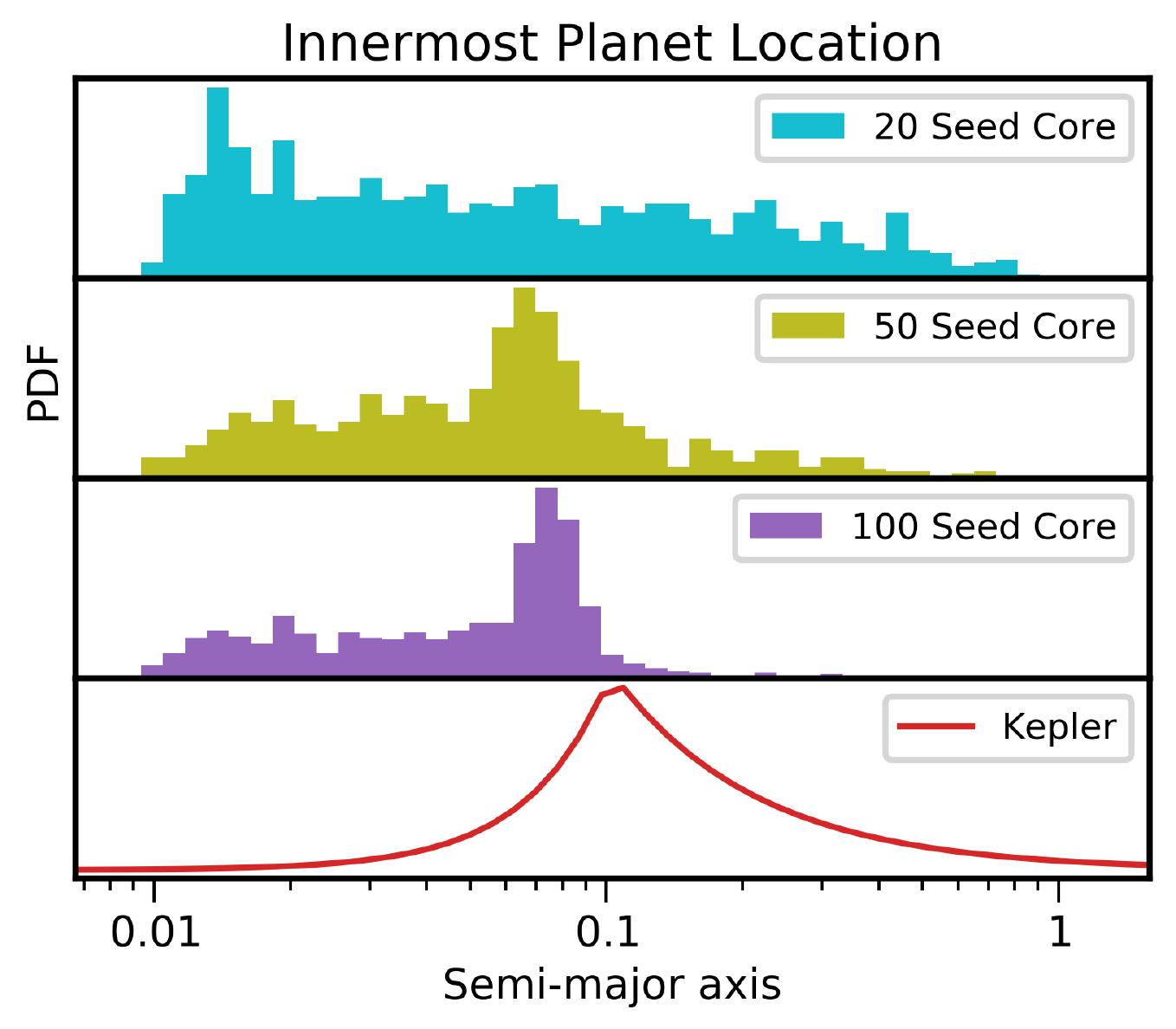}
    \includegraphics[width=0.45\linewidth]{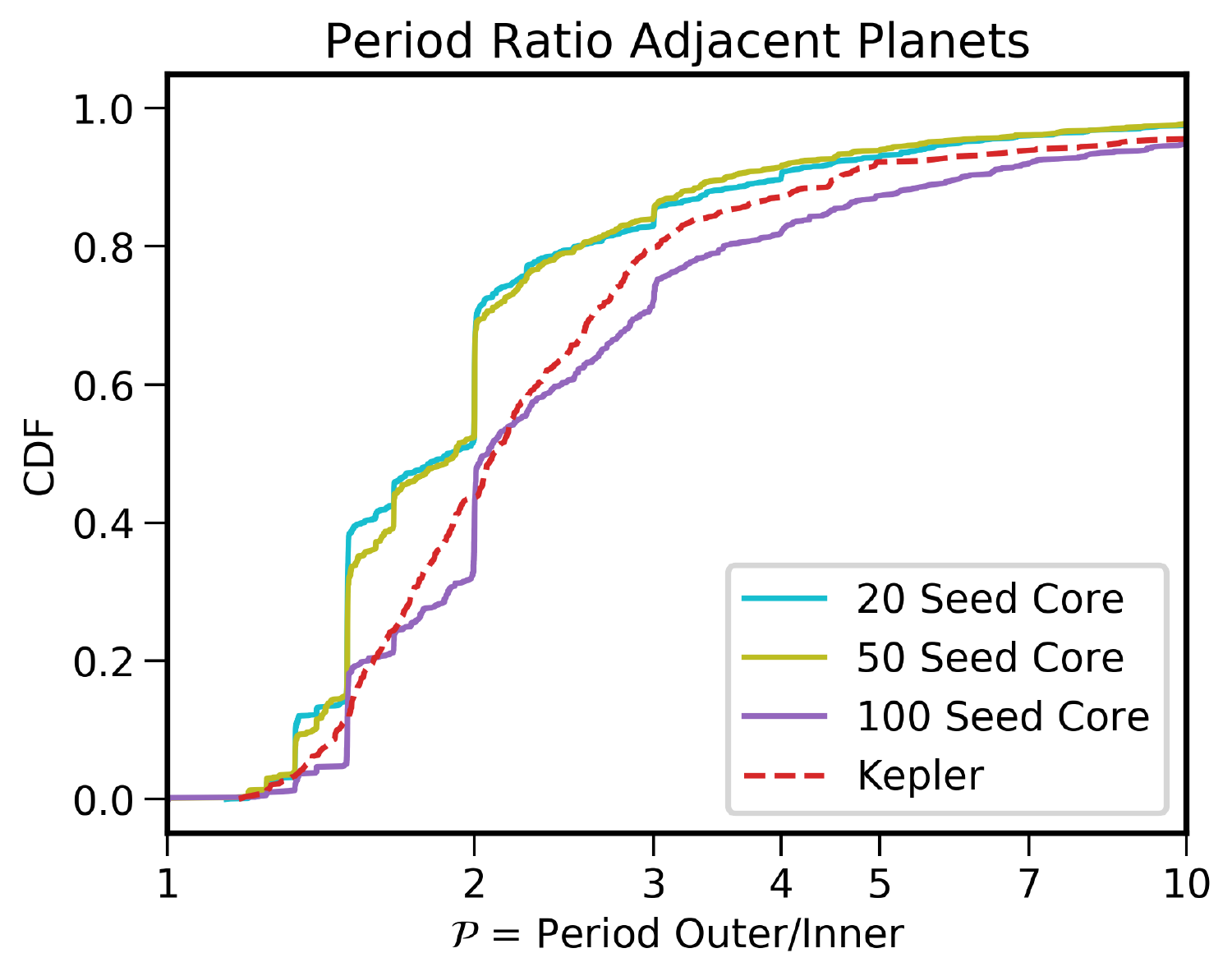}
    \setlabelscale
    \putlabel[2]{0.75}{a}
    \putlabel{0.75}{b}
     \caption{Location of the innermost planet in each planetary systems from the \bern models compared to those of \kepler.
     \Panel{a} shows the intrinsic distribution predicted by the model compared to the parametric distribution derived from \kepler.
     \Panel{b} shows the cumulative distribution in a simulated survey compared to the distribution observed with \kepler.
     \textbf{The innermost planets in each planetary system are located to close to the star, indicating the trap for migrating planets needs to be moved outwards.}
    }
    \label{f:all_Pin}
\end{figure*}

\subsection{Innermost Planet Location}\label{s:inner}
The location of the innermost planet in a planetary system traces the location of the gas disk inner disk edge --- which in turn is set by magnetospheric truncation --- and how efficiently migrating planets are trapped there.
The efficiency at which planets are pushed out of the trap by other planets in the system determines the distribution interior to the planet trap \citep[e.g.][]{2019MNRAS.tmp..963C}.
The distribution of the innermost planet in each system for the \bern models is shown in \sfigure{all_Pin}{a}. The inner planet distributions in the 50 and 100 seed core models show peaks just short-wards of $\sim 0.1$ au. The 100 core model shows a long tail towards smaller star-planet separations, while the 50 seed-core model also displays a tail toward longer periods, more similar to what is inferred from \kepler. The distribution of the innermost planet in the 20-core model does not show a clear peak but a wide distribution instead.
The peaks in the location of the innermost planet in the 50 and 100 core model nearly line up with the inferred peak from the \kepler data \citepalias{2018AJ....156...24M}, which is at $0.1$ au or $P=12$ days. However, the simulated distributions are skewed in a way that it over-predicts the number of planets at periods less than $10$ days and under-predicts the number of innermost planets at longer orbital periods. 

The overabundance of short-period planets in the 50 \bern models is most apparent from the simulated observations.  \sfigure{all_Pin}{b} shows the distribution of the innermost \textit{observed} planet in each planetary system in the simulated survey compared to the \kepler data. 
The simulated distributions peak around an orbital period of less than $2$ days, whereas the observed distribution peaks at longer periods ($P\approx 7$ days). This comparison highlights that the number of detected planets at short orbital periods is very sensitive to the location where planets are trapped. Because the transit probability increases rapidly with decreasing orbital period, the number of detectable planets increases to a level where it completely dominates the observational comparison if the occurrence rate at short orbital periods is high. 

Combined with the over-abundance of all planets at short orbital periods \sfigp{occ}{a}, this suggest that the migration trap at the disk inner edge needs to be placed at a larger distance from the star. The inner edge of the gas disk in the current models is placed at 0.03 au and the innermost planetesimals are located at twice that distance. A planet migrating alone can migrate down to the inner edge of the computational gas disk at 0.03 au. Planets even further in get pushed there by other planets when migrating within a resonant convoy of multiple planets.
A disk inner edge truncated at 0.1 au would likely provide a better match between the \bern models and the \kepler exoplanet population. This would facilitate a more in-depth comparison of the orbital period distribution interior to the disk inner edge to constrain the physics of migration and trapping.
A disk inner edge at 0.1 au is consistent with observations of the dust and gas in protoplanetary disks \citep[e.g.][]{2011ApJ...743..112S,2008ApJ...673L..63P}.

The distribution of orbital periods is generally broad. Although the distribution does not have the exact same shape as inferred from \kepler, it is not clear that a range of disk inner edge locations, as suggested by \cite{2017ApJ...842...40L} is needed to explain the observations. Instead, the stochastic nature of dynamical interactions during planet formation may be responsible for creating the broad period distribution interior to the disk inner edge (see also \citealt{2019MNRAS.tmp..963C})

\section{Predictions}\label{s:eta}
Population synthesis models make predictions for the occurrence and composition of planets even in regions that are not directly constrained by observations. Those predictions can now be informed and refined by empirical constraints from the observable planet population. Here, we briefly explore what the model that most closely matches the \kepler data predicts for 
the number and type of planets we expect to find in the habitable zone of a solar-mass star. While the \kepler stars span a range of stellar masses, the average stellar properties are representative of a sun-like star, and we leave an exploration of the spectral type dependence of habitable zone planets for a future paper. 

\begin{figure} 
    \includegraphics[width=\linewidth]{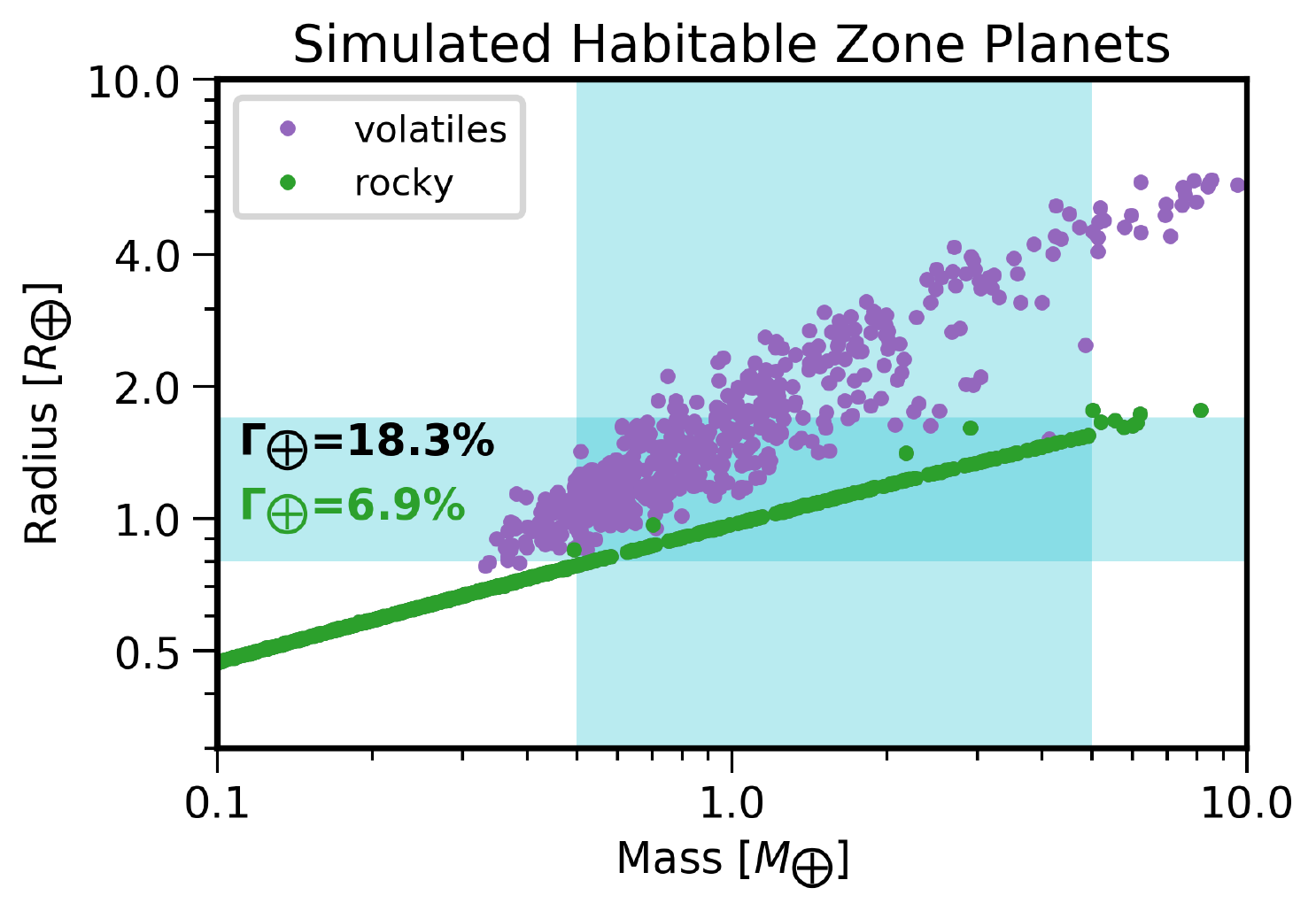}    
    \caption{Mass-radius relation of simulated planets in the conservative habitable zone \citep{2013ApJ...765..131K}. The dimensionless planet occurrence rate of earth-sized planets, $\Gamma_\oplus= \frac{d^2 N}{d \ln R\, d \ln P}$, is $\Gamma_\oplus=18\%$ for all planets but only $7\%$ when only rocky planets are included.
    }
    \label{f:eta}
\end{figure}

\rfigure{eta} shows the simulated masses and radii of planets in the habitable zone of the 100 seed-core model. The habitable zone is here defined as an orbital period range of $338-788$ days, corresponding to the conservative habitable zone \citep{2013ApJ...765..131K} around a sun-like star.
The populations that match the overall \kepler statistics, i.e. $\eta=0.2$ stars have a planetary system from the simulated set, predicts an occurrence rate of $0.8-1.7\,R_\oplus$ planets of $\Gamma_\oplus= \frac{d^2 N}{d \ln R\, d \ln P}= 18\%$. This rate is lower than the extrapolated power-law distributions from \citetalias{2018AJ....156...24M} at $\Gamma_\oplus=53\%$, which we have addressed in \cite{2019ApJ...883L..15P}. 

More importantly, the model predicts which planets in this size range are rocky and which planets have substantial hydrogen atmospheres. The fraction of true rocky planets is only $\Gamma_\oplus=7\%$, though we caution that this number is likely to change as the models become better calibrated to the observed exoplanet populations. However, future direct imaging missions should contend with the possibility that a large fraction of observable exo-Earths could have substantial hydrogen/helium atmospheres.

\section{Results and Discussion}
We have shown the need to evaluate planet formation models in a framework that accounts for the current observational biases in exoplanets surveys. 
This can be achieved with \epos both through direct comparisons with parametric distributions of planet properties and occurrence rates, as well as through performing simulated observations of synthetic planet populations. 

The \bern planet population synthesis models makes quantitative predictions for the distribution of planet and planetary system properties that we have evaluated with \epos using the \kepler exoplanet statistics. We generally find good agreement between the models and data over a large range of planetary (system) properties. The \bern population synthesis models with multiple (20, 50, 100) interacting seed planet cores per protoplanetary disk form co-planar planetary systems of sub-Neptunes at short orbital periods that are typical of the \kepler ``super-earths''. In contrast, synthetic populations with one seed planet core per disk produces many sharp features in the planet period-radius diagram that are not seen in exoplanet surveys.
The smooth observed distributions indicate that planets rarely form via the growth of isolated oligarchs, but instead grow through the interactions of multiple protoplanetary bodies.

A quantitative comparison also reveals a number of key differences that indicates parameters and processes that could be refined in the model to provide a better match to the \kepler database. These are: increasing the formation efficiency of mini-Neptunes, increasing the amount of gravitational interactions between proto-planets, moving the planet trap at the inner disk edge outward, and decreasing the overall occurrence of planets.

\subsection{Formation efficiency of mini-Neptunes}
Simulated surveys of the \bern models where 1 in 5 of stars have planetary systems drawn from the synthetic population provides roughly the right proportion of detectable Hot Jupiters ($\sim1\%$), Warm Giants ($\sim10\%$), and Hot Earths ($\sim50\%$). 
The \bern models under-predict the occurrence rate of mini-Neptunes ($\sim10\%$ simulated vs $\sim60\%$ observed) compared to rocky planets and warm giants. This indicates that a larger fraction of planetary cores need to accrete gaseous envelopes than in the current models, but without triggering more runaway growth into giant planets. 
A better understanding of gas accretion is needed and a more efficient formation channel for planets with substantial gaseous envelopes has to be incorporated in the synthesis model to match the \kepler planets. 
A similar trend is seen in micro-lensing surveys that probe larger planet-star separations, where the \bern model under-predicts the occurrence of intermediate-mass planets \citep{2018ApJ...869L..34S}. 

\subsection{Dynamical excitation of planetary orbits}
While the period ratio distribution of the 100 seed core model match closely with the observed distribution, too many planets are in orbital resonances. This may be indicative of a lack of dynamical interactions between planets in the simulations, which may in turn be a result of the short (20 Myr) N-body integration time considered here. Systems could go unstable at a later phase, as was seen in \cite{2017MNRAS.470.1750I}, and running the simulations longer could result in a better match to the observed systems.
Alternatively, external perturbations from stellar or planetary companions \citep[e.g.][]{2009MNRAS.394L..26M,2017MNRAS.467.1531H} or a reduction in the amount of gas to reduce orbital damping \citep{2016ApJ...822...54D} may be necessary components for a planet population synthesis model. 

\subsection{Trapping planets at the inner disk edge}
The \bern model over-predicts the occurrence of planets at short orbital periods, and the innermost planets of multi-planet systems are located too close to their host stars by a factor four. Future simulations could utilize a larger inner disk gas radius of $0.1$ au where planets are trapped, which would still yield planets inside of that location due to dynamical scattering. Such a radius would be consistent with observations of protoplanetary disks \citep{2011ApJ...743..112S} and coincide with the break in exoplanet occurrence rates that was previously inferred to be connected with the gas disk inner edge \citep{2015ApJ...798..112M,2017ApJ...842...40L}.

\subsection{Planet occurrence rates}
The \bern models predict a larger fraction of stars with planets than observed. The \kepler planet occurrence rates can be matched when only 1 in 5 of stars have planetary systems consistent with the simulated systems. A similar conclusion was reached by \cite{2019ApJ...874...81F} based on a comparison with more distant giant planets from radial-velocity surveys. 
A lower planet formation efficiency would provide a better match to the observed populations, for example if 4 in 5 stars form planets below the detection limits.
Such a low planet formation efficiency could be achieved by extending the distribution of protoplanetary disks to include lower mass disk that form planets below the detection limit, or by lowering the formation efficiency of planetesimals such that fewer solids end up in planets.
Alternatively, external factors such as binarity \citep{2016AJ....152....8K} or cluster interactions \citep{2018MNRAS.474.5114C} could reduce the fraction of stars with planets.

\section{Future Outlook}
The observational comparisons in this paper provide a means of properly evaluating how well synthetic planet populations match the planetary systems that have been detected to date. By understanding which diagnostics are matched between model and real systems, we are able to constrain the physical processes that occur during planet formation and calibrate the nuisance parameters used in such models. Iterating between these approaches allows us to make, with greater confidence, more quantitative predictions regarding planetary properties, such as bulk or atmospheric compositions, that will be measured by future observational facilities such as JWST and the next generation of telescopes.

\acknowledgments
We acknowledge a constructive and positive report from an anonymous referee. We would like to thank Bertram Bitsch, Daniel Angerhausen, Rachel Fernandes, Leslie Rogers, and Aaron Hamann for providing feedback on the manuscript and code. 
This material is based upon work supported by the National Aeronautics and Space Administration under Agreement No. NNX15AD94G for the program Earths in Other Solar Systems. The results reported herein benefited from collaborations and/or information exchange within NASAs Nexus for Exoplanet System Science (NExSS) research coordination network sponsored by NASA's Science Mission Directorate.
C.M. and A.E. acknowledge the support from the Swiss National Science Foundation under grant BSSGI0$\_$155816 ``PlanetsInTime''. Parts of this work have been carried out within the framework of the NCCR PlanetS supported by the Swiss National Science Foundation.\\

\software{
NumPy \citep{numpy}
SciPy \citep{scipy}
Matplotlib \citep{pyplot}
Astropy \citep{astropy}
\texttt{epos} \citep{epos}
KeplerPORTs \citep{2017ksci.rept...19B}
}

\ifhasbib
	\bibliography{popsynth.bbl}
\else	
	\bibliography{papers3,books,software}
\fi

\appendix
\section{Summary Statistics}

\begin{figure*} 
    \includegraphics[width=0.45\linewidth]{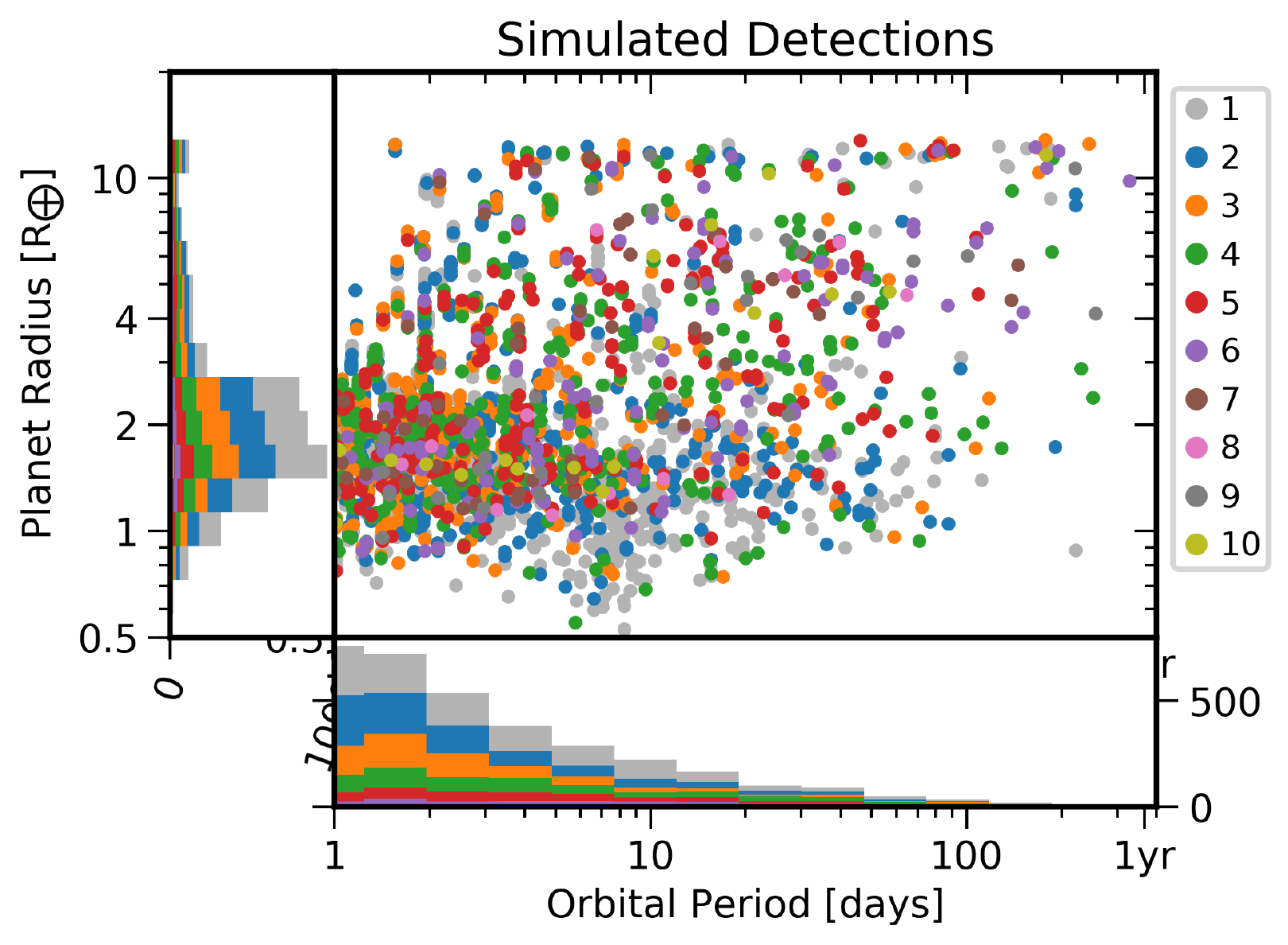}
    \includegraphics[width=0.45\linewidth]{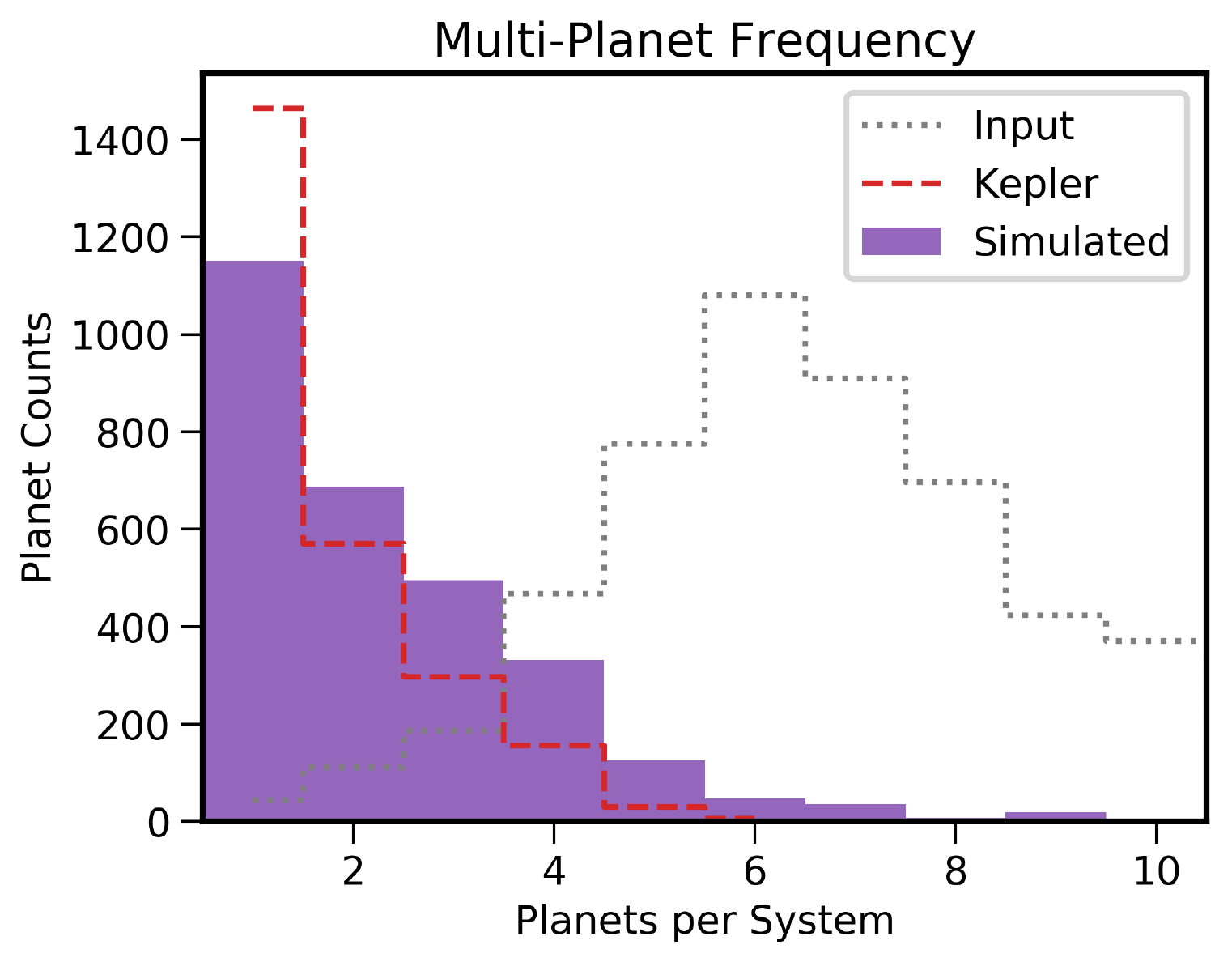}
    \setlabelscale
    \putlabel[2.0]{0.75}{a}
    \putlabel{0.75}{b}\\
    \includegraphics[width=0.45\linewidth]{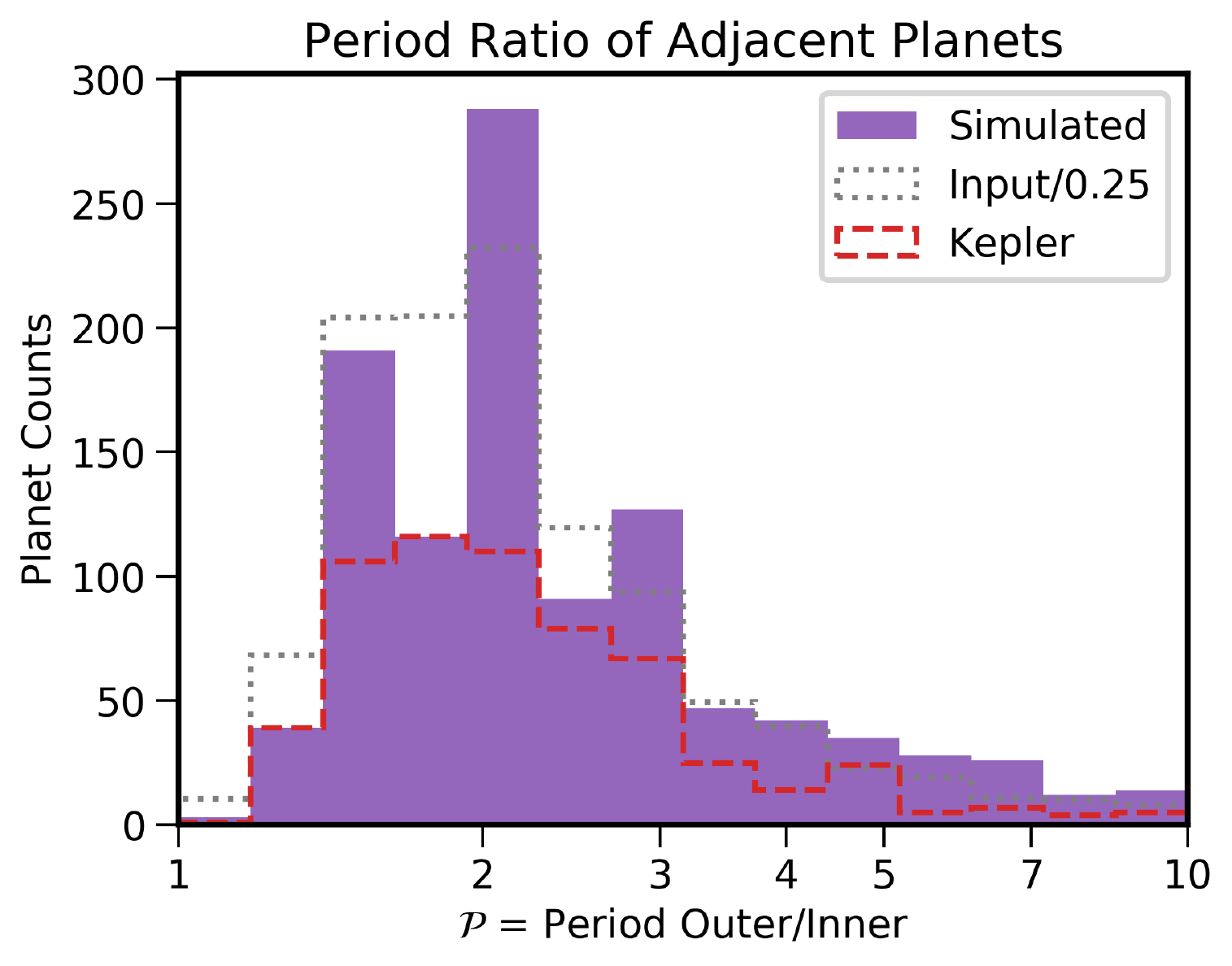}
    \includegraphics[width=0.45\linewidth]{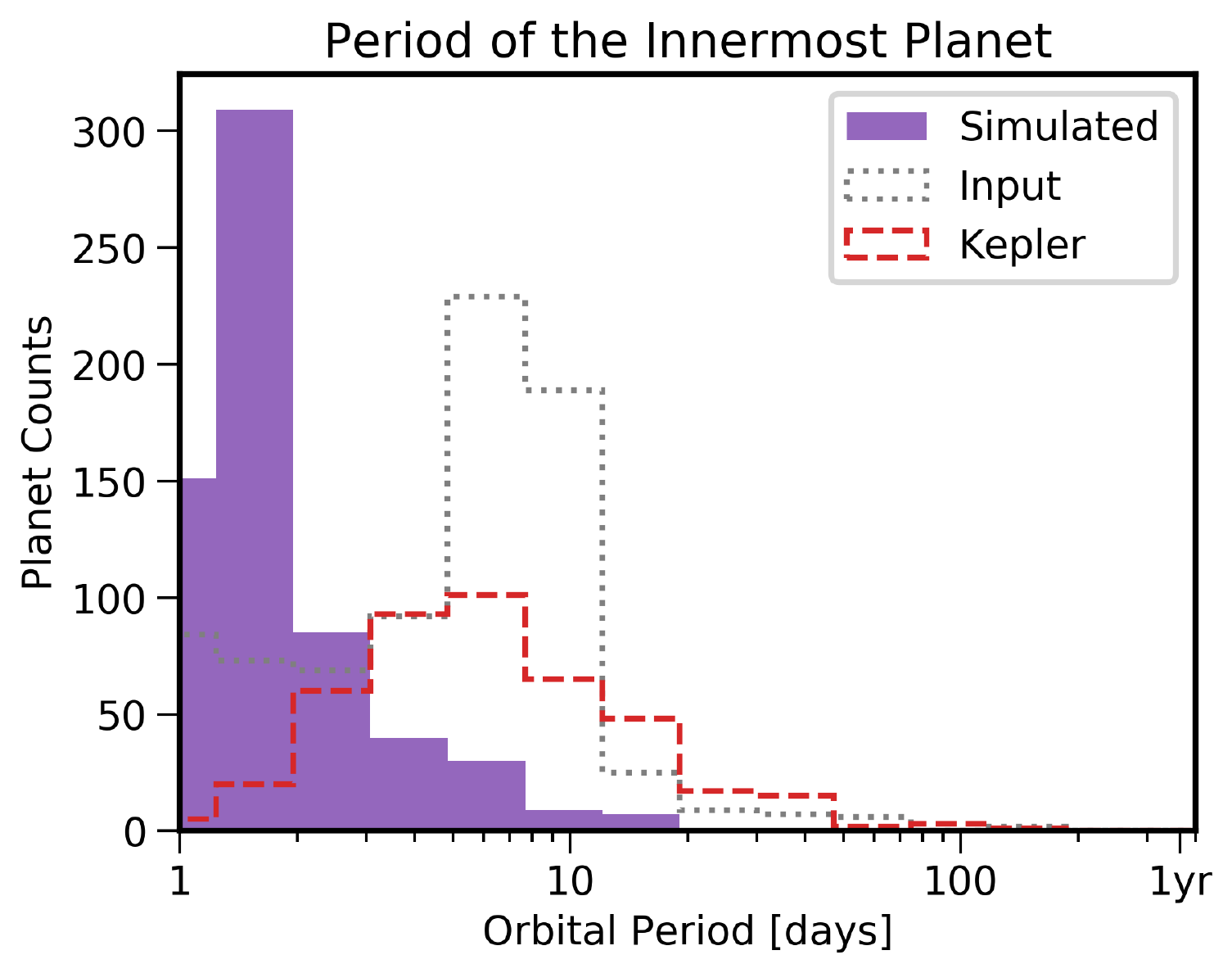}
    \putlabel[2.0]{0.75}{c}
    \putlabel{0.75}{d}
     \caption{\epos simulated observations of the \bern planet population synthesis model with 100 seed cores \panelp{a} and the associated summary statistics of multi-planet systems \panelp{bcd}. 
     \Panel{a} shows the orbital periods and planets radii of detectable planets, color-coded by number of observable planets in the system. 
     The summary statistics in panels \panelp{bcd} show the intrinsic distribution of planetary system properties from the \bern models with gray dotted lines, the observable distribution in the simulated survey in purple, and the distribution observed with \kepler in red dashed lines. All summary statistics are calculated for planets with orbital periods between $1$--$400$ days and radii between $0.5$--$14 ~R_\oplus$. 
     \Panel{b} shows the frequency of multi-planet systems, a tracer for planet inclination. 
     \Panel{c} show distribution of period ratios between adjacent planet pairs.
     \Panel{d} shows the distribution of the innermost planet in each multi-planet system.
    }
    \label{f:multi}
\end{figure*}

\rfigure{multi} show the summary statistics of the simulated populations of the 100 seed core \bern model compared with the equivalent observables from \kepler. These statistics are calculated for planets with sizes between $0.5$ and $20\,R_\oplus$ and orbital periods between $1$ and $400$ days. 

The multi-planet frequency --- the number of detectable planets per star --- of the 100 seed-core model is shown in \sfigure{multi}{b}. This statistic mainly traces the mutual inclination distribution but is also sensitive to planet spacings, sizes, and orbital periods. Most notably this statistic does *not* reflect the intrinsic multi-planet frequency (gray dotted line) as typically only a few planets in each multi-planet system are transiting, even in systems with low mutual inclinations. 

\sfigure{multi}{b} shows the distribution of period ratios between adjacent observed planets. The distribution before applying detection biases (gray line) peaks at shorter orbital periods than the observable one (purple). Generally, debiasing shifts the intrinsic period ratio distribution to larger values because planet pairs with a planet at large orbital periods are less likely to be observed \citep[e.g.][]{2016ApJ...821...47B}. The reason why we observe the opposite trend is the covariance between orbital period ratio and orbital period that can be seen in \sfigure{bern:multi}{b}: simulated planets at larger distances from the star have smaller period ratios. Because those planets also have low detection probabilities, they drop out of the observable sample in higher proportions than short-period, large period ratio planets, biasing the distribution.

\end{document}